\documentclass{article}

\usepackage{arxiv}

\usepackage[utf8]{inputenc} 
\usepackage[T1]{fontenc}    
\usepackage{url}            
\usepackage{booktabs}       
\usepackage{amsfonts}       
\usepackage{nicefrac}       
\usepackage{microtype}      
\usepackage{lipsum}		
\usepackage{graphicx}
\usepackage{doi}

\newtheorem{Numerical fact}[theorem]{Numerical fact}

\usepackage{algorithm}
\usepackage{algorithmic}
\usepackage{epstopdf}
\usepackage[caption=false]{subfig}
\usepackage{graphicx}
\usepackage{appendix}

\usepackage{makecell}
\usepackage{multirow}
\usepackage{amsmath}
\usepackage{array}
\usepackage{enumerate}
\usepackage{xcolor,soul}

\usepackage{subfig}
\usepackage{caption}

\title{Modeling for Non-exponential Production Systems Using Parts Flow Data: Model Parameter Estimation and Performance Analysis}

\author{ \href{https://orcid.org/0000-0002-6402-1711}{\includegraphics[scale=0.06]{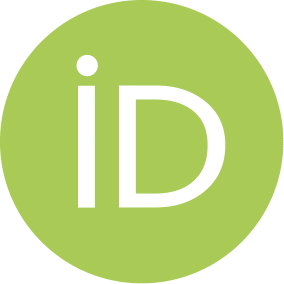}\hspace{1mm}Yuting Sun}\\
	Department of Electrical and Computer Engineering\\
        University of Connecticut \\
        Storrs, CT, USA\\
        \texttt{yuting.2.sun@uconn.edu} \\
	\And
	\href{https://orcid.org/0000-0002-3258-9636}{\includegraphics[scale=0.06]{orcid-1.png}\hspace{1mm}Liang Zhang} \\
	Department of Electrical and Computer Engineering\\
        University of Connecticut \\
        Storrs, CT, USA\\
        \texttt{liang.zhang@uconn.edu} \\
}

\renewcommand{\shorttitle}{\textit{arXiv} Template}

\begin{document}
\maketitle

\begin{abstract}
Mathematical modeling of production systems is the foundation of all model-based approaches for production system analysis, design, improvement, and control. 
To construct such a model for the stochastic process of the production system more efficiently, a new modeling approach has been proposed that reversely identifies the model parameters using system performance metrics (e.g., production rate, work-in-process, etc.) derived from the parts flow data. 
This paper extends this performance metrics-based modeling approach to non-exponential serial production lines. 
Since no analytical expressions of performance metrics are available for non-exponential systems, we use neural network surrogate models to calculate those performance metrics as functions in terms of the system parameters. 
Then, based on the surrogate models and given performance metrics, the machine parameters are estimated by solving a constrained optimization problem that minimizes the mean square error of the performance metrics resulting from the estimated parameters compared to the true ones.  
With the designed multi-start particle swarm optimization algorithm, we find that multiple non-unique combinations of machine parameters can lead to practically the same system performance metrics and a linear relationship of the reliability parameters from these obtained estimations is observed. Besides, model sensitivity analysis is 
implemented to verify the robustness of the different combinations of machine parameters even under the potential improvement scenarios.	
\end{abstract}


\section{Introduction}
\label{sec_intro}

In manufacturing research and practice, high-fidelity mathematical models are essential to implement model-based analysis, improvement, and control for production systems. 
Generally, the first step of production system modeling is to transform the system layout to a standard topological/structural model (\cite{cox1990mathematical, Meerkov:09}).
The basic standard entity of a production system includes machines and buffers. The machines can be individual equipment/workstations or a group of processing units such as human operators,
machines, cells, etc., and the buffers are the material handling devices, such as boxes, shelves, carts, conveyors, automated, guided vehicles, etc. Thus, a structural model is created by appropriately mapping each component of the real system into a standard entity in production system models, which is usually straightforward and with few variations in the result. 
Then, the next step is collecting factory floor data for each group of processing units and the material handling devices to identify the model parameters of machines and buffers. Unlike identifying the buffer capacities, which can be exactly measured, 
determining the reliability models of the machines poses one of the main challenges. On the one hand, the type of reliability models (e.g., Bernoulli, geometric, exponential, non-exponential, etc.) must be appropriate, i.e., should be convenient to analyze but not too simplified to lose the validation. On the other hand, the corresponding model parameters must lead to the practically same system performance metrics as the trues.   

During the past several decades, theoretical studies of production systems with different types of machine reliability models 
have accumulated a great number of results, and in most of the studies, the Markovian models, either discrete (i.e., Bernoulli and geometric models) or continuous (i.e., exponential model) are commonly used. 
Note that the Bernoulli reliability model is a special
case of the geometric reliability model where the sum of the breakdown rate and repair rate is 1. In the continuous-time cases, for the exponential reliability model, the up- and downtime of the machine are assumed to be exponential random variables, with the parameters, breakdown rate and repair rate, respectively. Both geometrical and exponential reliability models are characterized by Markov chains due to constant breakdown and repair rates. Clearly, the analytical methods are available to these Markovian models, which makes it easier to implement the performance evaluation and analysis. 
However, the up- and downtime of machines may follow non-Markovian or even arbitrary distributions. Moreover, as empirical evidence indicates, the coefficients of variation (CV) of up- and downtime of machines on the factory floor are often less than 1, so the corresponding probability distribution cannot be exponential. Thus, in some cases, the exponential reliability model cannot fit the real system as well as a non-exponential one. Unfortunately, 
since there are no analytical methods for performance analysis of non-exponential systems, even for the simplest two-machine cases, non-exponential reliability models are applied in very limited numbers of production system studies, even though they may lead to higher fidelity of the systems.  

Therefore, in this paper, we intend to deeply discuss the modeling approach and model performance for non-exponential production systems.   
Instead of the conventional modeling approach in which the operation status data is collected to identify the system mathematical model, 
the parts flow data-based modeling approach is applied. Specifically,
we measure the \textit{parts flow}, i.e., the entrances/exits of parts to/from the buffer and the number of parts in the buffer in each time slot, which are based on part-counting, 
and then, we develop an efficient algorithm to estimate the machine parameters of the non-exponential reliability models using the system performance metrics derived from the parts flow data as input.  
With the identified system models, we implement the model validity and sensitivity analysis. Additionally, the properties of the machine parameters and performance metrics are investigated.  

The rest of the paper is organized as follows: Section \ref{sec_review} reviews the mathematical models and analysis of non-exponential production systems and compares the typical approaches for conventional and new modeling approaches in the literature. 
Section \ref{sec_model} overviews the system model and the problems addressed in this paper. 
Mathematical modeling, especially, the algorithm of model parameter identification is developed in Section \ref{sec_modeling}. 
In Section \ref{sec_num}, model performance analysis is implemented and illustrated. Then,
the sensitivity of the model is discussed in Section \ref{sec_val}. 
Finally, the conclusions and future work are summarized in Section \ref{sec_con}.

\section{Literature Review}
\label{sec_review}

\subsection{Mathematical models and analysis for non-exponential production systems}

In production systems research, a production system is typically modeled as a stochastic process, where the operations of the machines are characterized by randomly distributed uptimes, downtimes, and/or cycle times (see \cite{Papdopoulos:93,Gershwin:94,Yao:94,Meerkov:09}).
The most commonly used mathematical models for characterizing such random behavior of production operations include the Bernoulli, geometric, and exponential reliability models, which are characterized by Markov chains and analytical methods are available to calculate the system performance metrics (throughput, work-in-process, etc.), such as 
\cite{jia2015finite,ju2016selective,feng2018analysis,jia2019serial,bai2021new}, etc.

Unfortunately, in practice, the machines often have up- and downtime distributed non-exponentially, and few results about the analytical calculation of performance metrics for production systems with non-exponential machines are available at this point. Among a limited number of studies, it is observed in \cite{Meerkov:09} that the steady-state throughput of non-exponential serial lines is approximately a linear function of the average CV of the machine up- and downtimes when all other system parameters remain fixed.
Similar results are reported in \cite{ching2008assembly} for assembly systems with non-exponential machines. However, other performance metrics and the joint effects of the mean and CV's of machine up- and downtimes are not discussed. Another research of assembly systems is implemented in 
\cite{chen2022analysis}, which introduces a system-theoretic approach  
to analyze the assembly-time performance (ATP) of assembly systems with collaborative robots. The assembly operations are described by stochastic processes with both individual (human operator and robot) preparation tasks and joint collaboration tasks, and gamma distributions are used to approximate task times and aggregate multiple interacting tasks to estimate ATP with the designed approximation method. 
Moreover, the paper \cite{kang2015analysis} studies the multi-product manufacturing systems with non-exponential processing times. Two approximation methods, gamma estimation and linear approximation, are proposed to estimate the system throughput. The model is validated with high accuracy by numerical experiments and practical data from an automotive assembly system.
In addition, the transient behavior of production serial lines with machines having gamma reliability models is investigated in \cite{yang2018transients}. However, only the cases in which all the machines have identical parameters are discussed.  
Besides, the paper \cite{kim2023throughput} suggests some different types of neural networks that approximate the throughput of Weibull serial production lines from given system parameters. Specifically, they focus on the cases where the $CV$'s for both uptimes and downtimes are greater than 1. 

\subsection{Conventional and new approach for parameter identification when modeling a production system}

In the conventional approach of production system modeling, we first determine the structural model based on the system layout, and then,  
usually collect the operating status data (i.e., up- and downtimes) from each individual workstation to identify the model parameters by  
designing customized procedures. The examples of conventional modeling processes are illustrated in the case studies of \cite{arinez2009quality, zandieh2017buffer, lee2018analysis, park2019improving}. 
The main limitations of this approach include high complexity and low quality of operating status data, which usually leads to significant efforts spent on collecting, cleaning, and processing these data. 
These limitations usually make it difficult to effectively apply the conventional approach to carry out modeling, analysis, and control for the production systems in practice.

To overcome these difficulties, a new modeling approach is proposed to reversely estimate the machine parameters of a production system model based on measured system performance metrics derived from the parts flow data. 
Two different types of methods have been developed to inversely estimate machine parameters in production serial line models: analytical expression-based method (\cite{sun2020parameter, sun2021parameter, sun2022novel, sun2022application}) and statistical/machine learning-based method (\cite{sun2020parameterber, tu2020estimation}). In both methods, standard system performance metrics (throughput, work-in-process, etc.) are used as the input to identify the machine parameters. 
In \cite{sun2020parameter}, based on the close-formed analytical expression of performance metrics and aggregation method, the problem of parameter identification for Bernoulli production serial line models is solved. Similarly, this method is also applied in synchronous and asynchronous exponential systems (\cite{sun2021parameter, sun2022novel, sun2022application}), and then, the applicability is verified in a practical industrial case study. On the other hand, \cite{sun2020parameterber} and \cite{tu2020estimation} apply the learning-based methods for parameter identification in Bernoulli and exponential production line models, respectively, in which the machine parameters are predicted by neural networks (and/or other learning models) with the performance metrics as input data. 
The main drawback of the learning-based method proposed in \cite{sun2020parameterber} and \cite{tu2020estimation} is that only one set of estimated machine parameters can be obtained from the learning models with given performance metrics data, so it is hard to determine whether multiple non-unique estimations exist.

\subsection{Research originality and contributions}

The main originality and contributions of the paper are as follows. 
\begin{itemize}
    \item The neural network-based surrogate models, i.e., the surrogate expressions of performance metrics in terms of system parameters are established for the multiple-machine non-exponential serial lines. 
    Instead of estimating the performance metrics by simulation, the surrogate models make the performance metrics estimation much more efficient.  

    \item An efficient and robust search algorithm for machine parameter estimation to match the given performance metrics is developed and multiple non-unique optimal solutions can be obtained.

    \item The accuracy of the estimated performance metrics resulting from the estimated model parameters obtained by the proposed algorithm is verified through numerical/statistical experiments. Moreover, the properties and numerical facts of the machine parameters and performance metrics are summarized. 
    
    \item The applicability of the identified models is demonstrated through model sensitivity analysis. Especially, the transformation between non-exponential and exponential models is discussed.
\end{itemize}

\section{System Description and Problems Addressed}
\label{sec_model}

\subsection{System model}
\begin{figure}[ht]
    \centering
    \includegraphics[trim= 10 400 300 0,clip,width=0.7\textwidth]{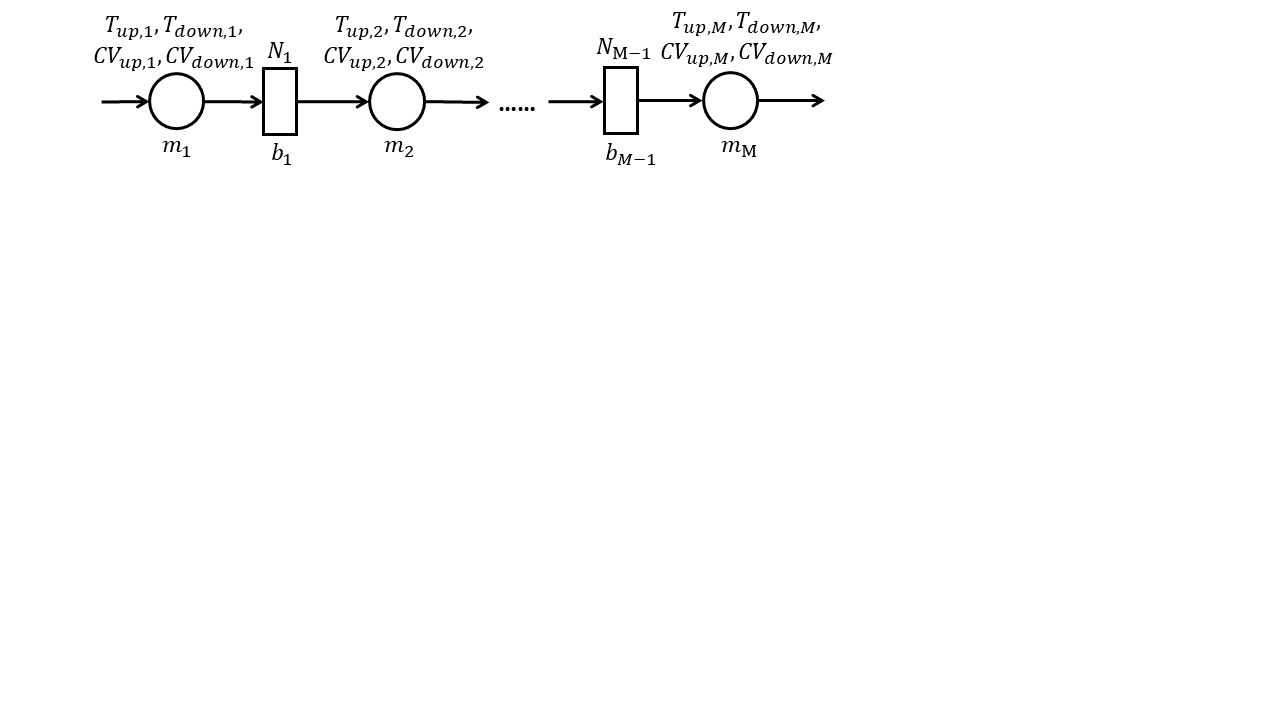}
    \caption{$M$-machine non-exponential serial line model}
    \label{fig_sys}
\end{figure}

In this paper, we study the synchronous $M$-machine serial production lines as shown in Figure \ref{fig_sys}, and the model assumptions are as follows.

\begin{enumerate}
\item All the machines have identical and constant cycle times. The model operates in continuous time and the time is measured in units of cycle time.
\item Machine $m_i$, $i=1,... M$, are unreliable and subject to random failures, i.e., the uptime and downtime of machine $m_i$ are random variables following distributions $f_{up,i}$ and $f_{down,i}$, respectively. The means and coefficient
of variations (CV's) of the up- and downtime of machine $m_i$ are denoted as $T_{up,i}$, $CV_{up,i}$ and $T_{down,i}$, $CV_{down,i}$, respectively.
The states (up and down) of the machines are independent. Using machine parameters $T_{up,i}$ and $T_{down,i}$, the efficiency of machine $m_i$, $e_i$, can be defined as 
\begin{equation} \label{e_f}
    e_i=\frac{T_{up,i}}{T_{up,i}+T_{down,i}}, \quad i=1,...,M.
\end{equation}

\item  Buffer $b_j$ has capacity $N_j$, i.e., it can store up to $N_j$ parts, $j=1,...,M-1$.
\item  Machine $m_i$ is starved if it is up, buffer $b_{i-1}$ is empty and machine $m_{i-1}$ is down. Machine $m_{1}$ is never starved. 
\item Machine $m_{i}$ is blocked if it is up, buffer $b_{i+1}$ is full, and machine $m_{i+1}$ is down. Machine $m_M$ is never blocked.
\item If machine $m_i$ is up and neither starved nor blocked, it processes parts with constant cycle time. 
\end{enumerate}

In practice, the distributions of up- and downtimes ($f_{up,i}$ and $f_{down,i}$) are typically unknown, but the system performances are insensitive to the particular $f_{up,i}$ and $f_{down,i}$ and depend mostly on their first two moments, which has been verified in \cite{Meerkov:09}. 
Among the different types of non-exponential distribution (e.g., gamma, Weibull, Log-normal, etc.), exponential distribution (which is very widely used as the machine reliability model) is a special case of gamma distribution ($CV_{up}=CV_{down}=1$), so in order to make the further investigation of the transformation between exponential and non-exponential models convenient, we assume that the up- and downtimes of all machines follow the gamma distribution.

\subsection{System performance metrics} \label{sec_spm}
Under system model assumptions, we define the following system performance metrics derived from the \textit{parts flow} data: 

\begin{itemize}
    \item \textit{Production rate}, $PR$: the average number of parts produced by $m_M$ per cycle time during steady state.
    \item \textit{Work-in-process}, $WIP_j$: the average number of parts contained in buffer $b_j$ during steady state.
    \item \textit{Probability that buffer $b_j$ is empty}, $P_{0,j}$: the fraction of time that there is no part in $b_j$ during steady state.
    \item \textit{Probability that buffer $b_j$ is full}, $P_{N,j}$: the fraction of time that there are $N_j$ parts in buffer $b_j$ during steady state.
    \item \textit{Probability of unchanged buffer state}, $B_{0,j}$: the probability that the number of parts in buffer $b_j$ is not changed compared with the last cycle time.    
\end{itemize}
In addition, we define four levels of buffer occupancy:
    \begin{align*}
        \text{Buffer occupancy level 1:}\; & \text{0 (not including 0) to 25\% of $N_j$};\\
        \text{Buffer occupancy level 2:}\; & \text{25\% of $N_j$ to 50\% of $N_j$};\\
        \text{Buffer occupancy level 3:}\; & \text{50\% of $N_j$ to 75\% of $N_j$};\\
        \text{Buffer occupancy level 4:}\; & \text{75\% of $N_j$ to $N_j$ (not including $N_j$)};
    \end{align*}
and then, we define
\begin{itemize}
\item \textit{Probability of buffer occupancy level $k$}, $P_{Lk,j}$: the fraction of time that the occupancy of buffer $b_j$ is at level $k$, $k=1,2,3,4$, during steady state. 
\end{itemize}
Thus, for each $b_j$, we have $P_{0,j} + P_{L1,j} + P_{L2,j} + P_{L3,j} + P_{L4,j} + P_{N,j} =1$. This implies that these six probabilities only have the degree of freedom equal to five.

\subsection{Problems addressed}
In this study, the following problems are addressed: 
\begin{itemize}
    \item \textbf{Model parameter identification}: Develop an efficient algorithm to identify the system model parameters (either unique or non-unique) using the performance metrics derived from the parts flow data. 
    \item \textbf{Model validation and performance analysis}: 
    Evaluate the efficiency of our proposed algorithm and implement the accuracy analysis of the performance metrics resulting from the estimated parameters. 
    \item \textbf{Model sensitivity analysis}: With the estimated parameters, verify they are still robust under buffer expansion and downtime reduction. 
\end{itemize}

\subsection{Challenges and proposed method}
\label{subs_appr}

Although the performance metrics' expression-based new modeling approach has been successfully applied to the systems with Bernoulli or exponential machines and proven high-fidelity system models can be obtained with this approach, however, a few challenges still exist for non-exponential production systems.      

\begin{itemize}
    \item The close-formed expressions of performance metrics of non-exponential systems are impossible to derive in an analytical way, even in the two-machine cases.  
    \item Although the performance metrics can be computed using simulation when the analytical expressions are not available, it takes so long that the computation time of iteratively searching for the optimal solution to match the performance metrics becomes unacceptable. 
    \item It is possible that there exist multiple (non-unique) solutions of estimated parameters, which can lead to the practically same performance metrics. Moreover, for some cases, the exponential fit also works well. There is no standard rule to select the model type and model parameters. 
    
\end{itemize}

\section{System model parameter identification}
\label{sec_modeling}
\subsection{Neural network surrogate models for performance metrics}
\begin{figure}[ht]
\centering
\includegraphics[trim= 0 110 385 15,clip,width=0.51\linewidth]{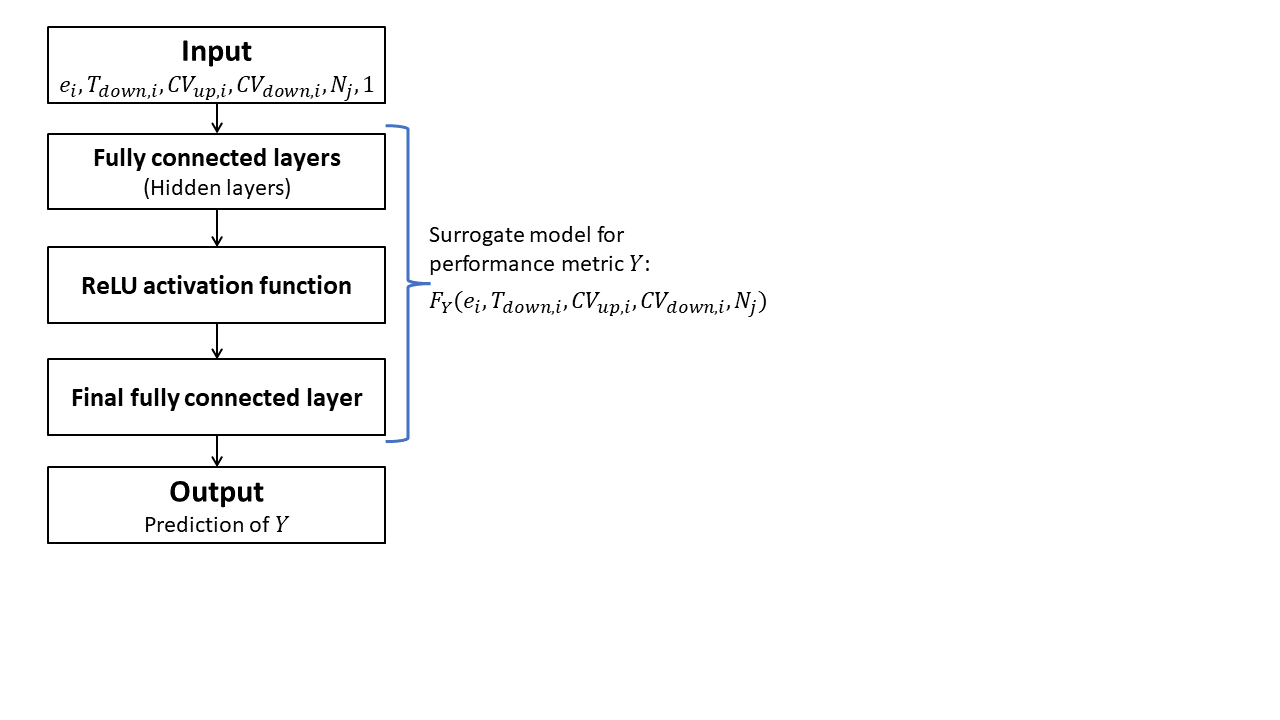}
\caption{Construction of the neural network for the surrogate model of each performance metric}
\label{fig-nn1}
\end{figure}

Since the analytical expressions of performance metrics of non-exponential systems are unavailable,
we first need to derive the quantitative relationships between the system parameters and performance metrics. In this paper, we propose to develop a surrogate model for calculating each performance metric as close-formed functions in terms of system parameters, $F_{Y}(e_i, T_{down,i}, CV_{up,i}, CV_{down,i}, N_j)$, $i=1,...,M$, $j=1,...,M-1$,
where $Y$ represents the system performance metrics defined in Section \ref{sec_spm}. 
In this study, the surrogate models are constructed using neural networks (NNs). The neural network is one of the most powerful models in machine learning. With the cooperation of interconnected units, called \textit{neurons}, a neural network model can learn the nonlinear relationship between the predictors and the responses very well. 
The construction of the neural network for the surrogate model of each performance metric is shown in Fig. \ref{fig-nn1}.

The neural network model for each performance metric consists of an input layer, several hidden layers, and an output layer following the activation function and the final fully connected layer. In the input layer, the number of nodes is set to the number of predictors plus one bias node. 
The number of neurons in the hidden layers depends on the complexity and the amount of training data. In the hidden layers, each neuron is connected to all neurons in the preceding layer by an associated numerical weight. The weight connecting two neurons regulates the magnitude of the signal that is transmitted between them. Finally, all neurons pass onto the ReLU activation function and the final fully connected layer, and then, a single node exports from the output layer, which is the numerical response, the prediction of performance metric $Y$  (e.g., $PR$, $WIP_j$'s, etc.). Note that, each performance metric for a system has an individual NN model. 
To train the NN models, we use limited-memory Broyden-Fletcher-Goldfarb-Shanno (LBFGS) quasi-Newton algorithm \cite{nocedal2006numerical} as optimizer.  In this study, since $1+7(M-1)$ performance metrics are considered, $1+7(M-1)$ independent NN models are developed for an $M$-machine serial production line. The numbers of neurons in hidden layers are selected as $(64,32)$ for all networks of three-machine cases and $(64, 32, 32)$ for all networks of five-machine cases.

\subsection{Algorithm to parameter identification}\label{para_id} 

With the Neural network surrogate models as the close-formed function to estimate the performance metrics, we propose to identify the machine parameters by minimizing the errors of estimated performance metrics compared with
the observed ones. We define the machine parameter in the form of $\mathbf{x}=(e_i, T_{down,i}, CV_{up,i}, CV_{down,i})$, $i=1,...,M$. 
Note that this formulation is equivalent to $(T_{up,i}, T_{down,i}, CV_{up,i}, CV_{down,i})$ and gives more convenience in determining the bounds of the machine parameters. 
Since the buffer capacities are easy to measure on the factory floor, we suppose $N_j$'s ($j=1,...,M-1$) are known, and then we define the vector-valued function $\mathbf{F}(\mathbf{x, N})$ as:
\begin{equation} \label{vf}
\begin{aligned}
&\mathbf{F}(\mathbf{x, N}) = \left [
         f_1(\mathbf{x, N}), ..., f_{K_{PM}}(\mathbf{x, N}) 
    \right ]^T \\
    &= \left [ \begin{array}{l}
         F_{PR}(e_i, T_{down,i}, CV_{up,i}, CV_{down,i}, N_j)-PR^*  \\
         F_{WIP_j}(e_i, T_{down,i}, CV_{up,i}, CV_{down,i}, N_j)/N_j-WIP_j^*/N_j \\
         F_{P_{0,j}}(e_i, T_{down,i}, CV_{up,i}, CV_{down,i}, N_j)-P_{0,j}^* \\
         F_{P_{N,j}}(e_i, T_{down,i}, CV_{up,i}, CV_{down,i}, N_j)-P_{N,j}^* \\
         F_{P_{L1,j}}(e_i, T_{down,i}, CV_{up,i}, CV_{down,i}, N_j)-P_{L1,j}^* \\
         F_{P_{L2,j}}(e_i, T_{down,i}, CV_{up,i}, CV_{down,i}, N_j)-P_{L2,j}^* \\
         F_{P_{L3,j}}(e_i, T_{down,i}, CV_{up,i}, CV_{down,i}, N_j)-P_{L3,j}^* \\
         F_{B_{0,j}}(e_i, T_{down,i}, CV_{up,i}, CV_{down,i}, N_j)-B_{0,j}^* \\
    \end{array}
    \right ]. 
\end{aligned} 
\end{equation}
where $ .^*$'s are the observed system performance metrics and $K_{PM}$ is the number of performance metrics we considered.  
Based on the above, we formulate the following constrained optimization problem: \vspace{5pt}\\
\textit{Find machine parameter $\mathbf{x}=(e_i, T_{down,i}, CV_{up,i}, CV_{down,i})$ that minimizes the 2-norm of error function $\mathbf{F}$ over a certain box-constraint set $\mathbf{X}$, i.e.,
\begin{align}
\begin{split}
    \min_{\mathbf{x}} & \; f(\mathbf{x}) = ||\mathbf{F}(\mathbf{x, N})||^2/K_{PM},\\
    \textrm{s.t.} & \; \mathbf{x} \in \mathbf{X},
\end{split}
\label{eq_org}
\end{align}
where \textit{where $\mathbf{X}=\{\mathbf{x}\in \mathbb{R}^{4M}|L_{k}< x_k< U_{k},k=1,...,4M\}$ and 
$L_{k}$, $U_{k}$ are the lower- and upper bounds for the parameters $x_k$'s.}
}
\vspace{5pt}

In this paper, to solve the parameter identification problem \eqref{eq_org} and to ensure global optimum, we thus develop the following multi-start particle swarm optimization (M-PSO) algorithm. 
In this algorithm, we start the PSO algorithm with $D$ different initialization sets in parallel. Since in the past study, we found that machine efficiencies can always be estimated with high accuracy using the PSO algorithm from any random initialization (see the justification of Numerical fact 1), we tighten the constraints of machine efficiencies using the first $D_n$ optimized solutions, i.e. we compute the average estimated machine efficiencies of first $D_n$ optimized solutions as $\Tilde{e}_i$, $i=1,...,M$ and let $L_i = \Tilde{e}_i-\Tilde{e}_i\cdot \epsilon_e\%$ and $U_i = \Tilde{e}_i+\Tilde{e}_i\cdot \epsilon_e\%$ as the new constraints for following $D-D_n$ runs of PSO. With the tightened constraints, the iteration may converge faster, and the quality of optimized solutions may improve.   
In summary, the procedure of the M-PSO algorithm is also described in the pseudo-code below.

\begin{algorithm}[H]
\small 
\caption{Multi-start Particle Swarm Optimization Algorithm (M-PSO)} 
\label{alg11}
\begin{algorithmic}
\FOR{$n=1,\dots, D$}
\STATE \textbf{Initialization}: Randomly create $K_p$ particles with the initial
particle position $\mathbf{x}_{(l)}^{(0)}\in \mathbf{X}$ and initial velocity $\mathbf{v}_{(l)}^{(0)}\in \mathbf{V}$, where $\mathbf{V}=\{\mathbf{v}\in \mathbb{R}^{4M}|L_{k}- U_{k} \le v_k\le U_{k}-L_{k},k=1,\dots,4M\}$, and $l=1,..., K_p$. \\
Set the stall counter $c = 0$, and iteration step $j=0$.
\STATE \textbf{Evaluation}: Find the best position 
$\mathbf{d}^{(0)}$ among $\mathbf{x}_{(l)}^{(0)}$ i.e., 
$\mathbf{d}^{(0)} = \arg \underset{\mathbf{x}_{(l)}^{(0)}} {\min}\; f(\mathbf{x}, N)$, and let $p_{(l)} = \mathbf{d}^{(0)}$, $l=1,\dots, K_p$.

\WHILE{$j<J_{max}$}
\FOR{$l=1,...,K_p$}
\STATE Randomly create neighborhood subset $S$ of $K_n$ particles other than $\mathbf{x}_{(l)}^{(j)}$ and find the best position $g_{(l)}$ among $S$;

\STATE Update the velocity: $\mathbf{v}^{(j+1)}_{(l)} = W\mathbf{v}^{(j)}_{(l)} + y_1 u_1\circ(p_{(l)}-\mathbf{x}_{(l)}^{(j)}) + y_2 u_2\circ(g_{(l)}-\mathbf{x}_{(l)}^{(j)})$, where $u_1$ and $u_2$ are randomly picked in $(0,1)$;

\STATE Update the position: $\mathbf{x}^{(j+1)}_{(l)} = \mathbf{x}^{(j)}_{(l)} +\mathbf{v}^{(j+1)}_{(l)}$;

\FOR{$k=1,...,4M$}
\IF{$x_k\notin \mathbf{X}_k$}
\STATE $x_k = \arg \underset{b} {\min}\; |b-x_k|$, $b=\{U_k, L_k\}$;
\IF{$v_k\notin \mathbf{V}_k$ } 
\STATE $v_k=0$;
\ENDIF
\ENDIF
\ENDFOR
\IF{$f(\mathbf{x}_{(l)}^{(j+1)}, \mathbf{N})<f(p_{(l)}, \mathbf{N})$}
\STATE $p_{(l)}=\mathbf{x}_{(l)}^{(j+1)}$;
\ENDIF
\ENDFOR
 \STATE $\mathbf{d}^{(j+1)} = \arg \underset{\mathbf{p}_{(l)}} {\min}\; f(\mathbf{p}_{(l)}, \mathbf{N})$;
\STATE $f_{best}^{(j+1)} = f(\mathbf{d}^{(j+1)}, \mathbf{N})$;
\IF{$f_{best}^{(j)}-f_{best}^{(j+1)} > 0$} \STATE $c=\max (0, c-1)$;
\IF{$c<2$} 
\STATE $W=\min (2W, U_w)$;
\ENDIF
\IF{$c>5$} 
\STATE $W=\max (W/2, L_w)$;
\ENDIF
\ELSE 
\STATE $c=c+1$ and $K_n = \min (K_n+K_n^{(0)}, K_p)$;
\ENDIF
\IF{ $|f_{best}^{(j)}-f_{best}^{(j+1)}| \geq \epsilon_f$ \AND $T<T_{max}$} 
\STATE $j=j+1$;
\ELSE 
\STATE Break;
\ENDIF
\ENDWHILE
\STATE $\hat{\mathbf{x}}_n = \mathbf{d}^{(j)}$.
\IF{$n=D_n$}
\STATE $\Tilde{e}_i = \frac{1}{D_n}\sum_{n=1}^{D_n} x_{i,n}$, $i=1,..,M$.
\STATE $U_i=\Tilde{e}_i+\Tilde{e}_i\cdot \epsilon_e\%$, $L_i=\Tilde{e}_i-\Tilde{e}_i\cdot \epsilon_e\%$, $i=1,..,M$.
\ENDIF
\ENDFOR
\STATE \textbf{Return} $\hat{\mathbf{x}}_1, \hat{\mathbf{x}}_2,..., \hat{\mathbf{x}}_D$.
\end{algorithmic}
\end{algorithm}

In the pseudo-code of this algorithm, we denote the number of particles as $K_p$, and the neighborhood size as $K_n$. The other hyperparameters include the initial inertia $W\in [L_w,U_w]$, self-adjustment weight $y_1\in(0,+\infty)$ and social adjustment weight $y_2\in(0,+\infty)$. Besides, we set the maximum number of iteration steps as $J_{max}$, and, if the running time (denoted as $T$) of solving a problem exceeds $T_{max}$, or the best value of the objective function cannot decrease $\epsilon_f$, then the algorithm is terminated.
Since multiple (non-unique) optimal solutions may exist, we start the algorithm with $D$ different sets of random initial points so that the distribution of optimized solutions can be further investigated.

\section{Numerical experiments and analysis}
\label{sec_num}
\subsection{Setup of numerical experiments}
\label{setup}
For the numerical experiments, a simulation program of the serial production line with gamma reliability machines is created that runs with the first 10,000 cycle times as warm-up time and with the next 300,000 cycle times being the time period to statistically evaluate the system performance metrics. Then, 40,000 different M-machine lines are randomly generated for each $M=\{3,5\}$. The system parameters are randomly selected from the following ranges: $e_i \in [0.7, 0.95]$, $T_{down,i}\in [3,20]$, $CV_{up,i}, CV_{down,i} \in [0.2, 1]$ and 
$\allowbreak N_j \in [\max\{T_{down,j}, T_{down,j+1}\},\; 3\cdot \max\{T_{down,j}, T_{down,j+1}\} ]$, for all $i=1,...,M$ and $j=1,...,M-1$.
For each line, 15 replications of the simulation program are executed and the average performance metrics from these runs are computed using the parts flow data collected from the buffer. All the computations are conducted in MATLAB on a Dell Inspiron 3671 workstation with Intel(R) Core(TM) i7-9700 CPU 3.00GHz processor and 16 GB of RAM.

\subsection{Model Validation and Performance Analysis}

\subsubsection{Accuracy of surrogate models}
For each $M=\{3,5\}$, among the 40,000 $M$-machine lines generated above, we randomly select 30,000 as the training dataset and the remaining 10,000 as the testing dataset. From the training data, we train the neural network model for each performance metric, i.e., with one of the performance metrics being the response and the system parameters ($e_i$'s, $T_{down,i}$'s, $CV_{up,i}$'s, $CV_{down,i}$'s and $N_j$, $i=1,...,M$, $j=1,...,M-1$) being the predictors. Then, we can obtain the NN surrogate model-based expression of each performance metric as a function of the system parameters, $F_{Y}(\mathbf{x},N)$, where $Y\in \{PR, WIP_j, P_{0,j}, P_{N,j} P_{L1,j}, P_{L2,j}, P_{L3,j}, B_{0,j}\}$. 
With the neural network surrogate models and given system parameters, we compute the performance metrics and evaluate the errors compared with the true ones using
\begin{equation}\label{err}
\begin{aligned}
     &\epsilon_{PR} = \frac{|\widehat{PR}-PR^*|}{PR^*}\cdot 100\%,\;\;\;\quad\quad
     \epsilon_{P_{0,j}} = |\widehat{P}_{0,j}-P_{0,j}^*|, \\
     &\epsilon_{WIP_{j}} = \frac{|\widehat{WIP}_{j}-WIP_{j}^*|}{N_j}\cdot 100\%, \;\; \epsilon_{P_{N,j}} = |\widehat{P}_{N,j}-P_{N,j}^*|, \\
     &\epsilon_{B_{0,j}} = \frac{|\widehat{B_{0,j}}-B_{0,j}^*|}{B_{0,j}^*}\cdot 100\%,\;\;\;\quad
     \epsilon_{P_{Lk,j}} = |\widehat{P}_{Lk,j}-P_{Lk,j}^*|,\; k=1,2,3. \\
\end{aligned}
\end{equation}
where $\widehat{\cdot}$ denotes the estimated performance metrics and $j=1,...,M-1$. 

Table \ref{tab_g35_rpm} shows the estimation errors of performance metrics of three-machine and five-machine gamma serial lines resulting from the neural network surrogate models. 
As one can see, given the system parameters, all performance metrics can be estimated with high accuracy using the surrogate models. 
This implies that our neural network surrogate models are sufficiently accurate for computing the performance metrics in such systems.
\begin{table}[ht]
\centering
\small
\caption{Estimation errors of NN surrogate models}
\label{tab_g35_rpm}
\begin{tabular}{lllll}
\hline \hline
     & \multicolumn{2}{c}{$M=3$}      & \multicolumn{2}{c}{$M=5$}      \\ \hline
     & Training data & Testing data & Training data & Testing data \\ \hline
$PR$   & 0.1311\%      & 0.1317\%     & 0.1923\%      & 0.2245\%     \\
$WIP_1$ & 0.4172\%      & 0.4477\%     & 0.4728\%      & 0.5913\%     \\
$WIP_2$ & 0.4164\%      & 0.4432\%     & 0.6752\%      & 0.8299\%     \\
$WIP_3$ &               &              & 0.6641\%      & 0.8327\%     \\
$WIP_4$ &               &              & 0.5026\%      & 0.6042\%     \\
$P_{0,1}$  & 0.0025        & 0.0027       & 0.0027        & 0.0036       \\
$P_{0,2}$  & 0.0027        & 0.0028       & 0.0050        & 0.0057       \\
$P_{0,3}$  &               &              & 0.0054        & 0.0063       \\
$P_{0,4}$  &               &              & 0.0038        & 0.0043       \\
$P_{N,1}$  & 0.0027        & 0.0028       & 0.0036        & 0.0040       \\
$P_{N,2}$  & 0.0027        & 0.0028       & 0.0055        & 0.0062       \\
$P_{N,3}$  &               &              & 0.0047        & 0.0058       \\
$P_{N,4}$  &               &              & 0.0030        & 0.0040       \\
$P_{L1,1}$ & 0.0025        & 0.0026       & 0.0027        & 0.0032       \\
$P_{L1,2}$ & 0.0028        & 0.0029       & 0.0039        & 0.0043       \\
$P_{L1,3}$ &               &              & 0.0045        & 0.0050       \\
$P_{L1,4}$ &               &              & 0.0041        & 0.0045       \\
$P_{L2,1}$ & 0.0026        & 0.0026       & 0.0038        & 0.0039       \\
$P_{L2,2}$ & 0.0025        & 0.0026       & 0.0043        & 0.0043       \\
$P_{L2,3}$ &               &              & 0.0048        & 0.0048       \\
$P_{L2,4}$ &               &              & 0.0046        & 0.0046       \\
$P_{L3,1}$ & 0.0027        & 0.0027       & 0.0042        & 0.0042       \\
$P_{L3,2}$ & 0.0024        & 0.0025       & 0.0046        & 0.0046       \\
$P_{L3,3}$ &               &              & 0.0044        & 0.0044       \\
$P_{L3,4}$ &               &              & 0.0032        & 0.0034       \\
$B_{0,1}$  & 0.2088\%      & 0.2139\%     & 0.2729\%      & 0.2980\%     \\
$B_{0,2}$  & 0.1908\%      & 0.1906\%     & 0.4219\%      & 0.4450\%     \\
$B_{0,3}$  &               &              & 0.5427\%      & 0.5581\%     \\
$B_{0,4}$  &               &              & 0.2765\%      & 0.3135\%    \\ \hline \hline
\end{tabular}
\end{table}

\subsubsection{Accuracy of system performance metrics resulting from estimated machine parameters}\label{res_b}
In this experiment, we randomly select 2,000 two-machine gamma lines from the testing dataset above. Given the true performance metrics as input, we search for the machine parameters using the M-PSO algorithm described in Section \ref{para_id} and the neural network surrogate models to calculate the performance metrics for each iterative solution found in the optimization process. For each line, we obtain $D=200$ optimized solutions from M-PSO with $K_p=200$ particles. For other hyperparameters of this algorithm, we set $K_n^{(0)}=25$, $W=U_w =1.1$, $L_w=0.1$, $y_1=y_2=1.5$, $\epsilon_f = 10^{-6}$, $J_{max} = 10000$, and $T_{max}=900s$.

Note that, while we obtain 200 different solutions of estimated machine parameters from M-PSO for each line, not all of them are necessarily \textit{valid}. Here, we define valid solutions $\hat{x}$ as those satisfying 
$\hat{x}\in \mathbf{X}$ and $f(\hat{x},N) < 10^{-4}$ (computed using either NN surrogate model or simulation). 
On average, for each three-machine and five-machine line, 155 and 168 out of 200 solutions found by M-PSO are valid, respectively.
Table \ref{tab_pm_sm1} shows the average errors of performance metrics under the \textit{valid} solutions of machine parameters estimated by M-PSO algorithm. The errors are evaluated based on \eqref{err} using both the NN surrogate model and simulation. As one can see, the machine parameters obtained can indeed provide an almost perfect match to the observed performance metrics under true system parameters.

\begin{table}[ht]
\centering
\small
\caption{Average of performance metrics estimation errors resulting from the estimated parameters}
\label{tab_pm_sm1}
\begin{tabular}{lllll}
\hline \hline
     & \multicolumn{2}{c}{$M=3$}                 & \multicolumn{2}{c}{$M=5$}                 \\ \hline
     & NN-based Err. & Simulation-based Err.  & NN-based Err. & Simulation-based Err. \\ \hline
$PR$   & 0.1686\%    & 0.2426\%               & 0.2561\%      & 0.4058\%               \\
$WIP_1$ & 0.1074\%   & 0.4064\%               & 0.2029\%      & 0.5212\%                \\
$WIP_2$ & 0.1293\%   & 0.3801\%               & 0.1835\%      & 0.7298\%                 \\
$WIP_3$ &            &                        & 0.1831\%      & 0.6768\%                  \\
$WIP_4$ &            &                        & 0.2001\%      & 0.5094\%                 \\
$P_{0,1}$  & 0.0012         & 0.0029                 & 0.0012               & 0.0056      \\
$P_{0,2}$  & 0.0011         & 0.0031                 & 0.0013               & 0.0068      \\
$P_{0,3}$  &                &                        & 0.0013               & 0.0082       \\
$P_{0,4}$  &                &                        & 0.0010               & 0.0067       \\
$P_{N,1}$  & 0.0009         & 0.0035                 & 0.0012               & 0.0053        \\
$P_{N,2}$  & 0.0010         & 0.0031                 & 0.0013               & 0.0086       \\
$P_{N,3}$  &                &                        & 0.0014               & 0.0075       \\
$P_{N,4}$  &                &                        & 0.0012               & 0.0051       \\
$P_{L1,1}$ & 0.0013         & 0.0027                 & 0.0021               & 0.0031        \\
$P_{L1,2}$ & 0.0011         & 0.0026                 & 0.0019               & 0.0043        \\
$P_{L1,3}$ &                &                        & 0.0022               & 0.0045        \\
$P_{L1,4}$ &                &                        & 0.0019               & 0.0041         \\
$P_{L2,1}$ & 0.0013         & 0.0026                 & 0.0023               & 0.0032         \\
$P_{L2,2}$ & 0.0012         & 0.0029                 & 0.0024               & 0.0043        \\
$P_{L2,3}$ &                &                        & 0.0027               & 0.0042         \\
$P_{L2,4}$ &                &                        & 0.0023               & 0.0035        \\
$P_{L3,1}$ & 0.0012         & 0.0026                 & 0.0023               & 0.0032        \\
$P_{L3,2}$ & 0.0013         & 0.0026                 & 0.0027               & 0.0041         \\
$P_{L3,3}$ &                &                        & 0.0024               & 0.0033           \\
$P_{L3,4}$ &                &                        & 0.0022               & 0.0030          \\
$B_{0,1}$  & 0.1632\%       & 0.2968\%               & 0.1942\%             & 0.4736\%         \\
$B_{0,2}$  & 0.1752\%       & 0.3123\%               & 0.1953\%             & 0.4957\%    \\
$B_{0,3}$  &                &                        & 0.2047\%             & 0.6188\%           \\
$B_{0,4}$  &                &                        & 0.1881\%             & 0.4124\%         \\
$f_{obj}$  & $2.55\times 10^{-6}$   &  $1.68\times 10^{-5}$    & $7.58\times 10^{-6}$   &$4.79\times 10^{-5}$  \\ \hline \hline                
\end{tabular}
\end{table}

\subsubsection{Distribution of estimated machine parameters} \label{est_par}
Although the valid solutions of estimated machine parameters found by M-PSO can fit the system performance metrics with very high accuracy, the underlying machine parameters may be quite different from the true parameters that we intend to identify.
However, M-PSO estimated parameters $\hat{T}_{down,i}$, $\widehat{CV}_{up,i}$, and $\widehat{CV}_{down,i}$ may distribute all over their feasible ranges defined in formulation \eqref{eq_org}.
Indeed, for all valid solutions of estimated machine parameters, the estimated machine efficiencies are typically very close to the true ones with low average estimation errors shown as Table \ref{tab_g_e}. Then, we obtain,

\begin{table}[ht]
\centering
\small
\caption{Estimation errors of machine efficiencies with/without tightened constraints}
\label{tab_g_e}
\begin{tabular}{llccccc}
\hline\hline
                        &       & $\epsilon_{e_1}$     & $\epsilon_{e_2}$     & $\epsilon_{e_3}$     & $\epsilon_{e_4}$     & $\epsilon_{e_5}$     \\ \hline
\multirow{2}{*}{$D_n$=20}  & $M=3$ & 0.1910\%             & 0.2553\%             & 0.1861\%             &                      &                      \\
                        & $M=5$ & 0.2872\%             & 0.3269\%             & 0.3319\%             & 0.3328\%             & 0.2531\%             \\ \hline
\multirow{2}{*}{$D_n$=200} & $M=3$ & 0.3432\% & 0.5034\% & 0.3333\%           &         &     \\
                           & $M=5$ & 0.7036\% & 0.6803\% & 0.6609\% & 0.6660\% & 0.6637\%  \\
                        \hline \hline

\end{tabular}
\end{table}

\textbf{Numerical Fact 1}: Given $N_j$'s and the system performance metrics set $\mathbf{Y}$, we can find multiple (non-unique) valid solutions for gamma parameters which can lead to practically the same performance metrics. For these estimated parameters, the solution of $\hat{e}_i$'s is unique and is the neighborhood of the true one. And the solutions of ($\hat{T}_{down,i}$, $\widehat{CV}_{up,i}$, $\widehat{CV}_{down,i}$) are non-unique. \\

\noindent\textit{Justification:} To justify this numerical fact, 2000 $M$-machine serial gamma lines are randomly selected for each $M=3,5$. For each case, we randomly select $D=200$ different initialization sets to start searching for the optimal solutions of estimated parameters using the M-PSO algorithm with $D_n = 200$ (without tightened constraints) or $D_n = 20$ (with tightened constraints), respectively. As a result, all the estimated machine efficiencies $\hat{e}_i$'s are very close to the true ones and the estimation errors are shown in Table \ref{tab_g_e}. This implies that the solution of $\hat{e}_i$'s is unique and is the neighborhood of the true one. Indeed, with tightened constraints, the accuracy of estimated machine efficiencies is significantly increased. Besides, for all the cases, there are many different valid solutions of  ($\widehat{T}_{down,i}$, $\widehat{CV}_{up,i}$, $\widehat{CV}_{down,i}$) distributed all over their feasible ranges.  
\vspace{5pt}

\begin{figure}[ht]
\centering
\includegraphics[trim= 50 10 40 10,clip, width=0.9\linewidth]{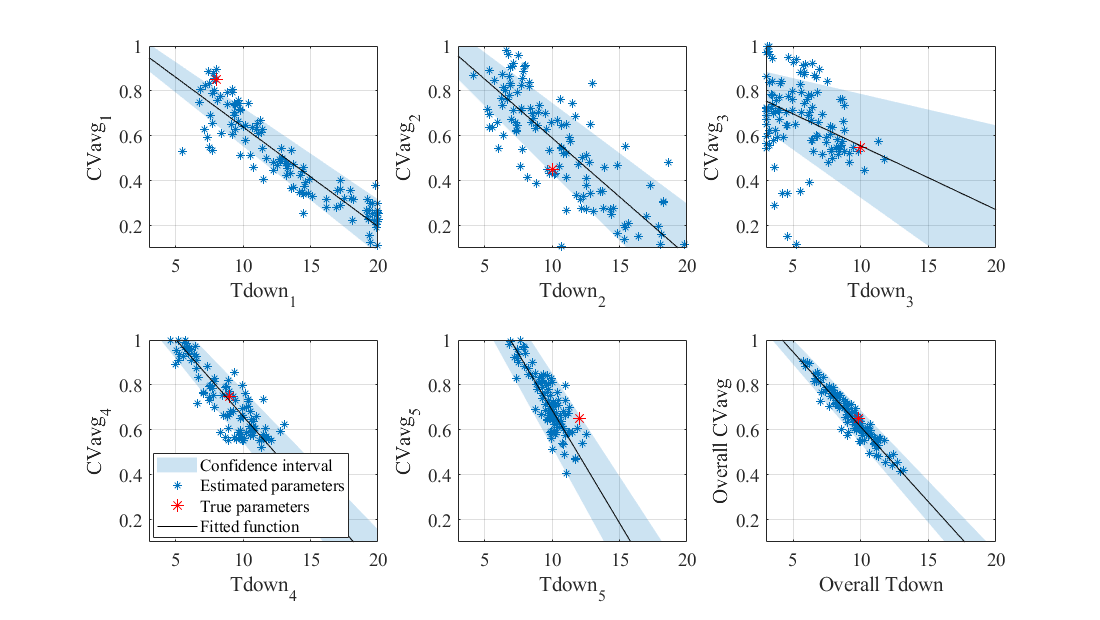}
\caption{$T_{down}$ vs. $CV_{avg}$ for estimated parameters of a five-machine case}
\label{fig1_g5c1_tcv}
\end{figure}

Since we find the estimated parameters $\hat{T}_{down,i}$, $\widehat{CV}_{up,i}$, and $\widehat{CV}_{down,i}$ may distribute all over their feasible ranges defined in formulation \eqref{eq_org}, the distribution of these estimated parameters is further investigated.
It is interesting to observe that the valid estimated machine parameters may exhibit a negative linear pattern between $CV_{avg}$ and $T_{down}$ for individual machines and the overall system. Taking a five-machine case (with the true parameters $e^*=(0.9, 0.85, 0.88, 0.9, 0.92)$, $T_{down}^*=(8,10,10,9,12)$, $CV_{up}^*=(0.8, 0.4, 0.5, 0.7, 0.6)$, $CV_{down}^*=(0.9, 0.5, 0.6, 0.8, 0.7)$, $N=(20, 15, 15, 20)$) as an example, this phenomenon is shown in Figure \ref{fig1_g5c1_tcv}. 
For individual machines, $CV_{avg,i} = \frac{1}{2}(CV_{up,i} + CV_{down,i})$. For the overall system, the overall downtime is defined as $\bar{T}_{down} = \frac{1}{M}\sum_{i=1}^{M} T_{down,i}$ and the overall CV is defined as $\bar{CV}_{avg} = \frac{1}{M}\sum_{i=1}^{M} \frac{1}{2}(CV_{up,i} + CV_{down,i})$. From Figure \ref{fig1_g5c1_tcv}, we can see the negative linear relationship between overall downtime and overall CV is very strong, but for the individual machine $m_3$, the linear relationship seems weak.

To evaluate the linear relationships, for each case, we create the linear regression models $F_i(T_{down,i}):\; CV_{avg,i} = b_{0,i} + b_{1,i} T_{down,i}$ for individual machines and $F_A(\bar{T}_{down}):\; \bar{CV}_{avg} = b_{0,a} + b_{1,a} \bar{T}_{down}$ for overall system.
Then, we compute the $p$-values of the coefficient $b_{1}$'s of these models. 
For the 2,000 cases of $M=\{3, 5\}$ in this experiment, we find significant linear relationships ($p$-value$<0.05$) between individual $T_{down,i}$ and $CV_{avg,i}$ (for all $i=1,...,M$) in 1521 and 1197 cases (out of 2,000). On the other hand, we find significant negative linear relationships between  $\bar{T}_{down}$ and $\bar{CV}_{avg}$ for all cases. Therefore, we obtain,

\textbf{Numerical Fact 2}: Given $N_j$'s and the system performance metrics set $Y$, we find multiple (non-unique) valid solutions for gamma parameters which can lead to practically the same performance metrics. Moreover, for these solutions, $\bar{T}_{down}$ and $\bar{CV}_{avg}$ have a significant negative linear relationship, and the fitted linear function $F_A(\bar{T}_{down}):\; \bar{CV}_{avg} = b_{0,a} + b_{1,a} \bar{T}_{down}$ can be found, in which the coefficients are significantly unequal to 0.  \\

\noindent\textit{Justification:} To justify this numerical fact, 2000 $M$-machine serial gamma lines are randomly selected for each $M=3,5$. The valid estimated machine parameters have been identified in Subsection \ref{res_b}. Then, we fit the linear regression model 
$F_A(\bar{T}_{down}):\; \bar{CV}_{avg} = b_{0,a} + b_{1,a} \bar{T}_{down}$ for each case. We compute the $p$-values of the coefficient $b_{1,a}$'s of these models. For all cases, we find $b_{1,a}<0$ and $p$-value$<0.05$, which implies $\bar{T}_{down}$ and $\bar{CV}_{avg}$ have a significant negative linear relationship.
Furthermore, this numerical fact is illustrated by the following groups of examples.
\vspace{5pt}

\textbf{Group 1}: All machines have identical parameters.
\begin{itemize}
    \item Group 1.1: $e_i=0.85$, $T_{down}=\{6, 9, 12\}$, $CV = \{0.3, 0.6, 0.9\}$, $N_j=15$.
    \item Group 1.2: $e_i=0.85$, $T_{down}=\{6, 12\}$, $CV = \{0.3, 0.6, 0.9\}$, $N_j=kT_{down}$, $k=2,3$.
\end{itemize}

\begin{figure}[p]
     \centering
     \subfloat[Overall $T_{down}$ vs. $CV_{avg}$ of estimated parameters for $M=3$ cases in Group 1.1]
         {\includegraphics[trim= 55 0 50 0,clip, width=1\linewidth]{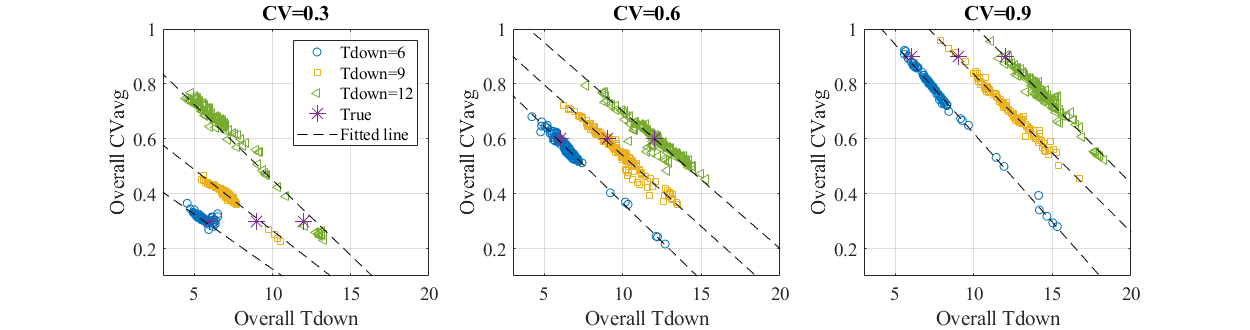}}
         \label{fig1_m3tcv1}\\
     \subfloat[Overall $T_{down}$ vs. $CV_{avg}$ of estimated parameters for $M=5$ cases in Group 1.1]
         {\includegraphics[trim= 55 0 50 0,clip, width=1\linewidth]{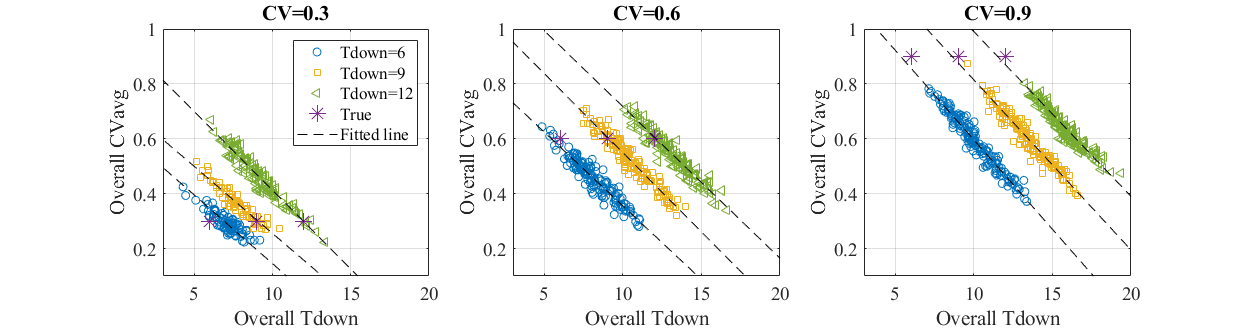}}
         \label{fig1_m5tcv2}\\         

     \subfloat[Overall $T_{down}$ vs. $CV_{avg}$ of estimated parameters for $M=3$ cases in Group 1.2] 
         {\includegraphics[trim= 90 0 80 0,clip, width=1\linewidth]{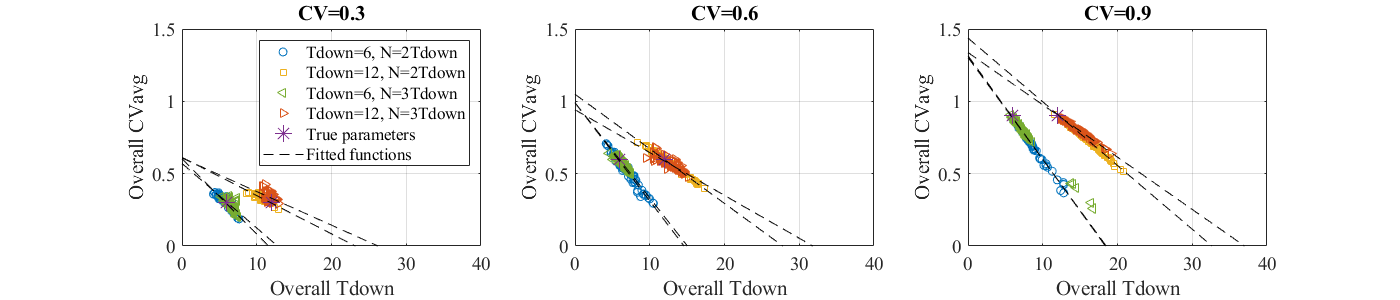}}
         \label{fig1_g3_n_cv}\\
     \subfloat[Overall $T_{down}$ vs. $CV_{avg}$ of estimated parameters for $M=5$ cases in Group 1.2] 
         {\includegraphics[trim= 90 0 80 0,clip, width=1\linewidth]{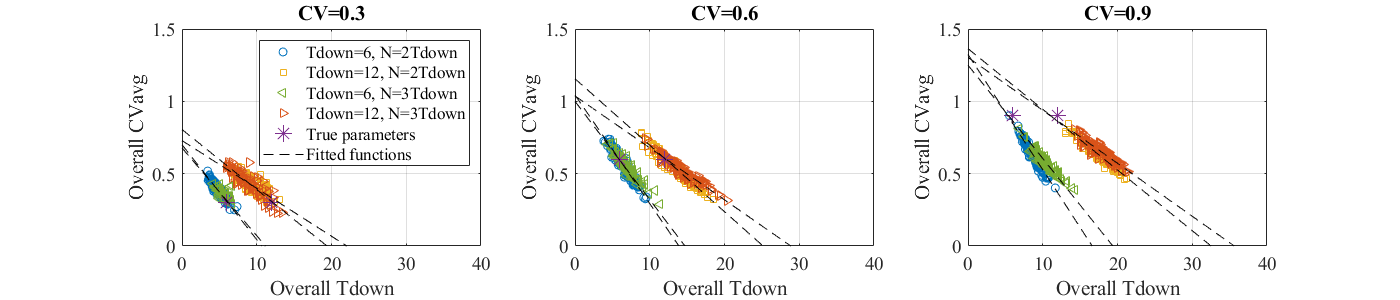}}
         \label{fig1_g5_n_cv}\\
         
        \caption{Overall $T_{down}$ vs. $CV_{avg}$ for estimated parameters of Group 1}
        \label{fig_grp1_cv}
\end{figure}

We plot the overall downtime and overall CV of the estimated parameters with the corresponding fitted linear functions for all the cases in this group as Figure \ref{fig_grp1_cv}. Note that, in the discussion of the slope of all fitted functions $F_A$, we compare the absolute values of $b_{1,a}$'s, so if an $F_A$ becomes steeper, we regard this as an increase of slope.
According to Figures \ref{fig_grp1_cv} (a) and (b) for Group 1.1 in which all buffer capacities are fixed as 15,  fitted linear functions $F_A$'s of the systems with the same CV are almost parallel, although the system overall downtimes are different. Moreover, the intercepts of  $F_A$'s increase with both overall $CV$ and $T_{down}$ of the systems. In addition, from Figures \ref{fig_grp1_cv} (c) and (d) for Group 1.2, when all the buffer capacities increase from two times to three times of the downtimes, this makes no significant effect on the fitted linear functions. Furthermore, if the systems have the same CV, the intercepts of $F_A$'s are almost the same, but the slopes of $F_A$'s decrease with downtimes of these systems. 
Similar to Figures \ref{fig_grp1_cv} (a) and (b), we can observe, if the systems have the same downtime, their corresponding intercepts and slopes of $F_A$'s both increase with system CV.

\vspace{5pt}
\allowdisplaybreaks
\textbf{Group 2}: All machines have different $T_{down}$'s and $CV$'s, but $\bar{T}_{down}$ or $\bar{CV}_{avg}$ are set as $\bar{T}_{down} = \{6,8,10,12\}$, $\bar{CV}_{avg}$=\{0.5, 0.75\}. 
Specifically, for $M=5$ cases as examples, $e=(0.85, 0.85, 0.85, 0.85, 0.85)$, $N=(15, 15, 15, 15)$, and
\begin{align*}
T_{down}1 =& \{(6,5,7,7,5), (6,8,7,4,5)\},\;(\bar{T}_{down}=6)\\ 
T_{down}2 =& \{(6,8,10,7,9), (6,12,9,6,7)\},\;(\bar{T}_{down}=8)\\ 
T_{down}3 =& \{(9,12,6,10,13), (9,7,12,10,12)\},\;(\bar{T}_{down}=10)\\ 
T_{down}4 =& \{(10,12,13,11,14), (13,9,12,14,12)\},\;(\bar{T}_{down}=12)\\
CV1 = &\{(0.5, 0.45, 0.4, 0.65, 0.55, 0.5, 0.4, 0.8, 0.4, 0.35),\\  
&(0.6, 0.5, 0.45, 0.55, 0.4, 0.35, 0.7, 0.55, 0.6, 0.3)\}, (\bar{CV}_{avg}=0.5)\\
CV2 = &\{(1, 0.75, 0.5, 0.8, 0.85, 0.95, 0.7, 0.9, 0.4, 0.65), \\
&(0.5, 0.9, 1, 0.65, 0.45, 0.85, 0.6, 0.75, 0.8, 1)\}, (\bar{CV}_{avg}=0.75)
\end{align*} 

We plot the overall downtime and overall CV of the estimated parameters with the corresponding fitted linear functions for the cases with all combinations of above $T_{down}$ and $CV$ in this group as Figure \ref{fig_avg_t_cv_m5}. The examples of $M=3$ cases are shown in Appendix A.2. In Figure \ref{fig_avg_t_cv_m5}, for instance, "CV1.1: 0.5" means the first assignment of CV1 is selected and the corresponding $\bar{CV}_{avg}$ is 0.5. The similar representation is also created for $T_{down}$, e.g., "Td1.1: 6" means the first assignment of $T_{down}$1 is selected and the corresponding $\bar{T}_{down}$ is 6.

\begin{figure}[H]
\centering
{{\includegraphics[trim= 0 0 15 0,clip, scale=0.5]{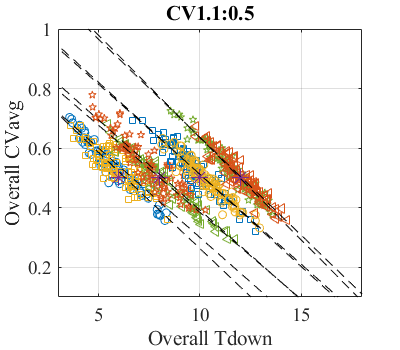}}
{\includegraphics[trim= 0 0 15 0,clip, scale=0.5]{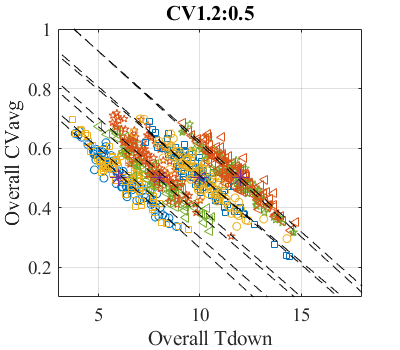}}\\
{\includegraphics[trim= 0 0 15 0,clip, scale=0.5]{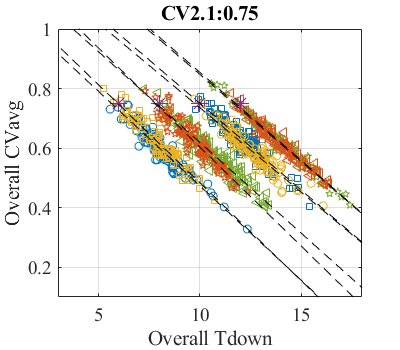}}
{\includegraphics[trim= 0 0 15 0,clip, scale=0.5]{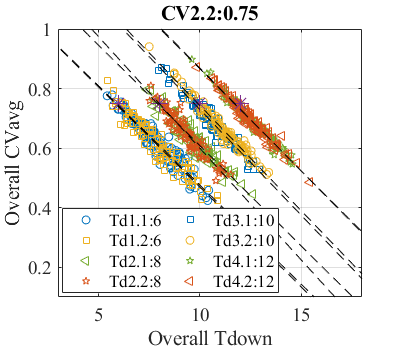}}}
\caption{Overall $T_{down}$ vs. overall $CV_{avg}$ of estimated parameters for $M=5$ cases in Group 2}
\label{fig_avg_t_cv_m5}
\end{figure}

According to Figure \ref{fig_avg_t_cv_m5} for Group 2, we can see, when the overall $T_{down}$ is fixed, under the same overall $CV_{avg}$, these different systems may share almost the same fitted linear functions of overall $T_{down}$ vs. overall $CV_{avg}$, although $T_{down,i}$'s, $CV_{up,i}$'s and $CV_{down,i}$'s are assigned with different ways. Moreover, if the systems have the same overall $T_{down}$ but different overall $CV_{avg}$, on average the slopes slightly increase with $CV_{avg}$, and the intercepts significantly increase with $CV$.
\vspace{5pt}

\textbf{Group 3}: All machines have different $T_{down}$'s, and $CV$'s are grouped by low, medium, and high levels. 
For $M=5$, we set $e=(0.85, 0.85, 0.85, 0.85, 0.85)$, $N=(15, 15, 15, 15)$, and
\begin{align*}
&T_{down}1=(6,8,7,10,9), \bar{T}_{down}=8\\
&T_{down}2=(8,12,10,6,9), \bar{T}_{down}=9\\
&T_{down}3=(8,6,10,9,12), \bar{T}_{down}=9\\
&T_{down}4=(9,7,12,10,12), \bar{T}_{down}=10
\end{align*}

Low CV: 
\begin{align*}
    &CV1 = (0.38, 0.30, 0.37, 0.41, 0.40, 0.30, 0.42, 0.45, 0.32, 0.35), \bar{CV}_{avg}=0.3700\\
    &CV2 = (0.46, 0.34, 0.36, 0.30, 0.39, 0.47, 0.43, 0.42, 0.41, 0.47), \bar{CV}_{avg}=0.4050\\
    &CV3 = (0.48, 0.50, 0.42, 0.48, 0.48, 0.47, 0.46, 0.45, 0.44, 0.30), \bar{CV}_{avg}=0.4480
\end{align*}

Medium CV: 
\begin{align*}
    &CV1 = (0.46, 0.77, 0.37, 0.87, 0.38, 0.54, 0.39, 0.52, 0.70, 0.40), \bar{CV}_{avg}=0.5400\\
    &CV2 = (0.52, 0.42, 0.51, 0.54, 0.79, 0.78, 0.55, 0.32, 0.79, 0.89), \bar{CV}_{avg}=0.6110\\
    &CV3 = (0.52, 0.76, 0.59, 0.42, 0.72, 0.65,  0.91, 0.82, 0.60, 0.76), \bar{CV}_{avg}=0.6750
\end{align*}

High CV: 
\begin{align*}
    &CV1 = (0.80, 0.82, 0.87, 0.80, 0.96, 0.80,    0.80, 0.79, 0.80, 0.86), \bar{CV}_{avg}=0.8300\\
    &CV2 = (0.75, 0.98, 0.98, 0.95, 0.77, 0.81,   0.83, 0.92, 0.78, 0.93), \bar{CV}_{avg}=0.8700\\
    &CV3 = (0.99, 0.98, 0.76, 0.94, 0.81, 0.85,   0.89, 0.99, 0.85, 1.00), \bar{CV}_{avg}=0.9060
\end{align*}

We plot the overall downtime and overall CV of the estimated parameters with the corresponding fitted linear functions for the cases with all combinations of above $T_{down}$ and $CV$ in this group as Figure \ref{fig_cv_m5}. The examples of $M=3$ cases are shown in Appendix A.3.

\begin{figure}[ht]
\centering
\subfloat[Overall $T_{down}$ vs. overall $CV_{avg}$ of estimated parameters for the low CV cases in Group 3]
{{\includegraphics[trim= 0 0 15 0,clip, scale=0.52]{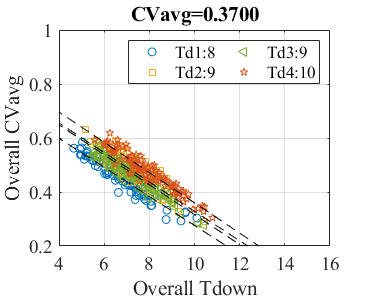}}
{\includegraphics[trim= 0 0 15 0,clip, scale=0.52]{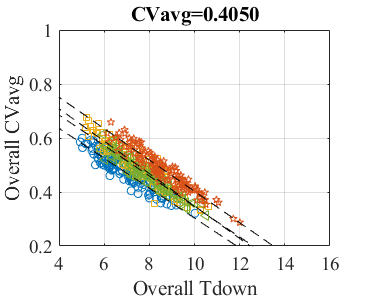}}
{\includegraphics[trim= 0 0 15 0,clip, scale=0.52]{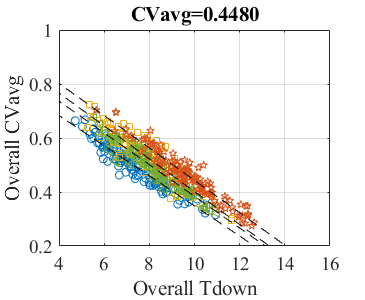}}
}
\label{fig_lcv1_m5}\\

\subfloat[Overall $T_{down}$ vs. overall $CV_{avg}$ of estimated parameters for the medium CV cases in Group 3]
{{\includegraphics[trim= 0 0 15 0,clip, scale=0.52]{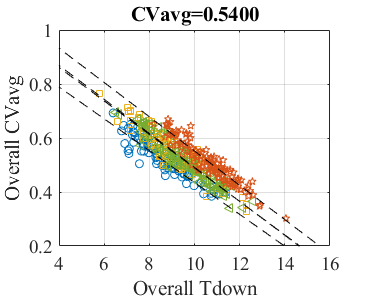}}
{\includegraphics[trim= 0 0 15 0,clip, scale=0.52]{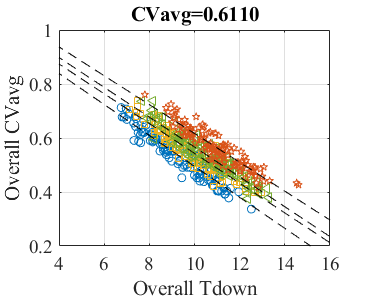}}
{\includegraphics[trim= 0 0 15 0,clip, scale=0.52]{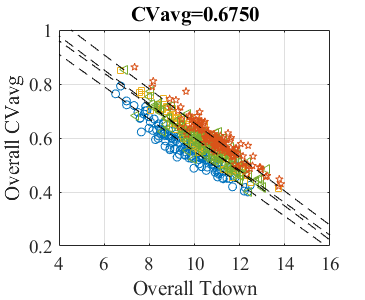}}
}
\label{fig_mcv1_m5}

\subfloat[Overall $T_{down}$ vs. overall $CV_{avg}$ of estimated parameters for the high CV cases in Group 3]
{{\includegraphics[trim= 0 0 15 0,clip, scale=0.52]{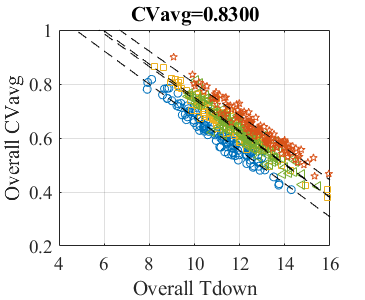}}
{\includegraphics[trim= 0 0 15 0,clip, scale=0.52]{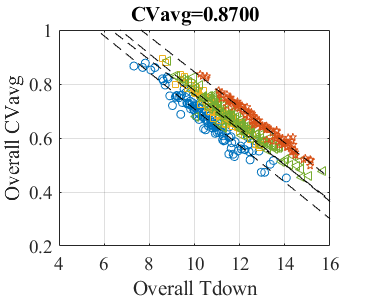}}
{\includegraphics[trim= 0 0 15 0,clip, scale=0.52]{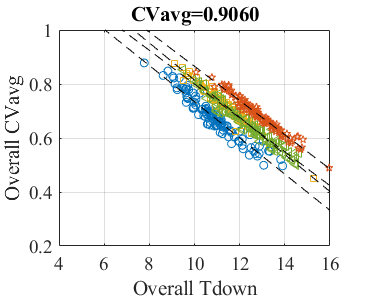}}
}
\label{fig_hcv1_m5}

\caption{Overall $T_{down}$ vs. overall $CV_{avg}$ of estimated parameters for $M=5$ cases in Group 3}
\label{fig_cv_m5}
\end{figure}

According to Figure \ref{fig_cv_m5} for Group 3, it is clear to find, again, the intercepts of fitted functions $F_A$'s we obtain from all valid estimated parameters increase with both overall $T_{down}$ and $CV_{avg}$ of these systems, and the fitted functions for those systems with the same overall CV but different overall downtime are almost parallel with each other. Besides, given $\bar{CV}_{avg}=1$, we can find the corresponding $\bar{T}_{down}$ based on $F_{A}$ fitted for this system. If $\bar{T}_{down}$ is not in a feasible range, e.g., $\bar{T}_{down}<0$, it implies we may not find a set of valid exponential parameters for this system. Otherwise, this system can be fitted as an exponential model, although the true CV of up- and downtimes are less than 1. From Figure \ref{fig_cv_m5}, clearly, the systems with higher CV are more possible to transform into the exponential model.

\vspace{5pt}
\textbf{Group 4}: All machines have different parameters and the orders of machines are assigned in different ways.
For $M=3$, the the buffer capacity is set as $(15, 20)$ and machine parameters 
for 3 machines are set as $m_i: (e, T_{down}, CV_{up}, CV_{down})$. Specifically,
\begin{equation}
    \begin{aligned}
        &m_a: (0.9, 10, 0.6, 0.4);\\
        &m_b: (0.85, 12, 0.9, 0.5);\\
        &m_c: (0.8, 8, 0.5, 0.8).
    \end{aligned}
\end{equation}
So we obtain the following lines featured as different machine efficiency patterns,
\begin{equation*}
    \begin{aligned}
        \text{Increasing pattern: } &m_c\rightarrow m_b\rightarrow m_a;\\
        \text{Invert bowl pattern: } &m_c\rightarrow m_a\rightarrow m_b;\\
        \text{Bowl pattern: } &m_a\rightarrow m_c\rightarrow m_b;\\
        \text{Decreasing pattern: } &m_a\rightarrow m_b\rightarrow m_c.
    \end{aligned}
\end{equation*}

For $M=5$, the the buffer capacity is set as $(15, 20, 15, 20)$ and machine parameters 
for 5 machines are set as $m_i: (e, T_{down}, CV_{up}, CV_{down})$. Specifically,
\begin{equation}
    \begin{aligned}
        &m_a: (0.95, 11, 0.8, 0.6);\\
        &m_b: (0.9, 10, 0.6, 0.4);\\
        &m_c: (0.85, 12, 0.9, 0.5);\\
        &m_d: (0.8, 8, 0.5, 0.8);\\
        &m_e: (0.75, 9, 0.4, 0.7);\\
    \end{aligned}
\end{equation}
Similarly, we obtain the following lines featured as different machine efficiency patterns,
\begin{equation*}
    \begin{aligned}
        \text{Increasing pattern: } &m_e\rightarrow m_d\rightarrow m_c\rightarrow m_b\rightarrow m_a;\\
        \text{Invert bowl pattern: } &m_e\rightarrow m_b\rightarrow m_a\rightarrow m_c\rightarrow m_d;\\
        \text{Bowl pattern: } &m_a\rightarrow m_c \rightarrow m_e \rightarrow m_d \rightarrow m_b;\\
        \text{Decreasing pattern: } &m_a\rightarrow m_b\rightarrow m_c \rightarrow m_d \rightarrow m_e;\\
        \text{Oscillating pattern: } &m_a\rightarrow m_c\rightarrow m_b \rightarrow m_e \rightarrow m_d;\\
    \end{aligned}
\end{equation*}

\begin{figure}[ht]
\centering
{\includegraphics[trim= 0 0 15 0,clip, scale=0.525]{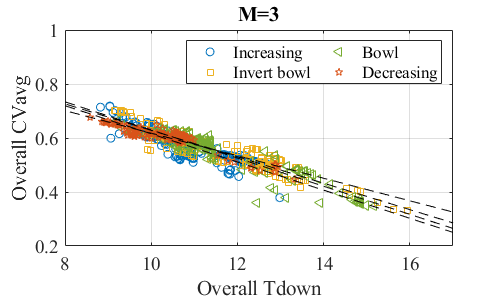}}
{\includegraphics[trim= 0 0 15 0,clip, scale=0.525]{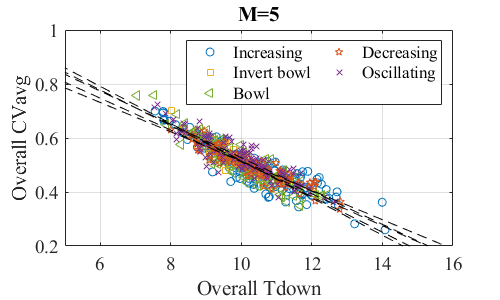}}
\caption{Overall $T_{down}$ vs. $CV_{avg}$ for estimated parameters of Group 4}
\label{fig_avg_e}
\end{figure}

Unlike Groups 1 to 3 in which we set all machine efficiencies as 0.85, the systems in Group 4 all have different machine efficiencies featured as several types of patterns.
According to Figure \ref{fig_avg_e}, we can see, for both three-machine and five-machine cases, the distributions of $(\bar{T}_{down}, \bar{CV}_{avg})$ of estimated parameters and their corresponding fitted functions are almost overlapped. This implies that machine efficiencies have little effect on the intercepts and slopes of the fitted linear functions when the different systems share the same $\bar{T}_{down}$ and $\Bar{CV}_{avg}$. 
\vspace{5pt}

According to the justification of Numerical Fact 2 and illustrative examples in 4 groups, we not only verify the negative linear relationship between $\bar{T}_{down}$ and $\bar{CV}_{avg}$ of the estimated parameters which all lead to practically the same performance metrics but also some properties of this linear relationship are observed.      
Furthermore, we then obtain,

\textbf{Numerical Fact 3}: Given $N_i$'s (with/without $e_i$'s or $\hat{e}_i$'s), and the fitted linear functions $F_A(\bar{T}_{down}):\; \bar{CV}_{avg} = b_{0,a} + b_{1,a} \bar{T}_{down}$, if we randomly select an average estimation point ($\bar{T}_{down}$, $\bar{CV}_{avg}$) on function $F_A$, then we can find at least a set of parameters of M machines ($T_{down,i}$, $CV_{up,i}$, $CV_{down,i}$, $i=1,...,M$) satisfying $F_A$ that lead to practically the same performance metrics.\\

\noindent\textit{Justification:} To justify this numerical fact, 500 out of 2000 $M$-machine serial gamma lines in testing data are randomly selected for each $M=3,5$. 
The valid estimated machine parameters have been identified in the above experiments, so $F_A(\bar{T}_{down})$ can be obtained. 
Next, we randomly select $\bar{T}_{down}$ and calculate the corresponding $\bar{CV}_{avg}$ based on $F_A(\bar{T}_{down})$. $T_{down,i}$, $CV_{up,i}$, $CV_{down,i}$ and $e_i$ are still all unknown variables but must match the average values of $\bar{T}_{down}$ and $\bar{CV}_{avg}$.
With the selected overall average constraints, we search for the best combination of $m_1$, $m_2$ and $m_3$ parameters that minimizes the performance metrics estimation error function $f(\hat{\mathbf{x}},N)$. 
For each case, 10 different average points are selected from $F_A$ and this experiment is conducted 10 times for each average point to obtain a total of 100 combinations of machine parameters. 
As a result, for each case under each average point, at least one machine parameter combination obtained above satisfies $f(\hat{\mathbf{x}},N)<10^{-4}$. In other words, for each line, the system performance metrics under all 10 different combinations of machine parameters studied are practically indistinguishable. 
\vspace{5pt}

Numerical Fact 3 also implies infinite numbers of non-unique solutions of estimated parameters, which lead to practically the same performance metrics as those observed ones, can be found. Moreover, given the observed performance metrics and  
a certain $\bar{T}_{down}$ (or $\bar{CV}_{avg}$) in a feasible range, we can always find a valid solution (or more than one) of estimated parameters. 
In practice, when we have experiences with approximated average downtime of all machines in a certain system (the overall CV usually is much harder to determine or estimate on the factory floor), then the corresponding estimation of all machine parameters can be obtained with the method described in Numerical Fact 3. Further discussion of the robustness of these estimated parameters is in the next section.

\subsection{Model Sensitivity Analysis}
\label{sec_val}

Although the model parameters identified can perfectly fit the observed system performance metrics, it does not directly imply that these model parameters are indeed the \textit{true} parameters of the production system. 
Thus, in this section, we investigate the sensitivity of the models identified through the proposed modeling approach.
Specifically, we first verify the robustness of multiple estimated gamma parameters (at least one of all $\widehat{CV}$'s $<1$) under improvement scenarios, e.g., buffer expansion, downtime reduction, etc. 
Then, we discuss the model type sensitivity, i.e., what if we fit a gamma system as an exponential model (all $\widehat{CV}$'s $=1$).   

\subsubsection{Sensitivity analysis of estimated gamma parameters} \label{par_sen_p}
According to numerical facts 2 and 3, we first find the fitted linear function for a case. Then, we randomly select several average estimation points and search for valid solutions with the average estimation constraints.

For instance, we take another two five-machine cases from Group 3 as examples, i.e., 
$e=(0.85, 0.85, 0.85, 0.85, 0.85)$, $N=(15, 15, 15, 15)$, $T_{down}=(8, 6, 10, 9, 12)$ ($\bar{T}^*_{avg}=9$)\\
Case 1: $CV = (0.99, 0.98, 0.76, 0.94, 0.81, 0.85, 0.89, 0.99, 0.85, 1.00)$, $\bar{CV}^*_{avg}=0.9060$\\
Case 2: 
$CV = (0.38, 0.30, 0.37, 0.41, 0.40, 0.30, 0.42, 0.45, 0.32, 0.35)$, $\bar{CV}^*_{avg}=0.3700$\\
The fitted linear functions of these 2 cases are\\
Case 1: $y=-0.0627x + 1.4261$\\
Case 2: $y=-0.0554x + 0.8700$\\
We select several average estimation points from the fitted linear function and obtain the following valid solutions in Table \ref{tab_est5_c12}. Additional examples of three-machine cases are shown in Appendix B. Since all the estimated machine efficiencies obtained for these cases are very close to the true ones (i.e., all $\hat{e}\approx 0.85$), they are not shown in Table \ref{tab_est5_c12}.

\begin{table}[ht]
\centering
\small
\caption{Selected estimations of $M=5$ cases 1 and 2}
\label{tab_est5_c12}
\begin{tabular}{llccl}
\hline\hline \\ [-0.75em]
&  & \multicolumn{1}{c}{$\bar{T}_{down}$}  & \multicolumn{1}{c} {$\bar{CV}_{avg}$} & \multicolumn{1}{c}{Solutions}\\  \hline \\[-0.75em]
\multirow{16}{*}{Case 1}&  &   &    & $\hat{\mathbf{T}}_{down} = (6.78, 5.69, 8.44, 8.98, 10.11)$ \\
&Est. 1  &8   &0.93    & $\widehat{\mathbf{CV}}_{up} = (1.00, 0.91, 1.00, 1.00, 1.00)$ \\
&     &    &        & $\widehat{\mathbf{CV}}_{down} = (1.00, 0.91, 1.00, 0.55, 0.88)$ \\ [+0.5em]
&   &   &   & $\hat{\mathbf{T}}_{down} = (8.55, 9.24, 11.06, 11.24, 9.91)$  \\
& Est. 2    &10    & 0.80 & $\widehat{\mathbf{CV}}_{up}= (0.95, 0.50, 0.29, 0.74, 0.99)$  \\
&     &     &       & $\widehat{\mathbf{CV}}_{down}= (0.80, 0.87, 0.96, 0.90, 1.00)$ \\ [+0.5em]
& &   &   & $\hat{\mathbf{T}}_{down} = (9.77, 8.95, 12.55, 15.16, 13.57)$  \\
& Est. 3    & 12  & 0.67  & $\widehat{\mathbf{CV}}_{up} = (0.64, 0.80, 0.40, 0.88, 0.38)$  \\
&     &     &     & $\widehat{\mathbf{CV}}_{down} = (0.95, 0.89, 0.49, 0.41, 0.90)$ \\ [+0.5em]
&    &   &    & $\hat{\mathbf{T}}_{down} = (6.68, 9.47, 8.75, 10.00, 10.10)$    \\
& Est. 4    & 9   & 0.86  & $\widehat{\mathbf{CV}}_{up} = (1.00, 0.62, 0.67, 1.00, 0.90)$   \\
&     &    &        & $\widehat{\mathbf{CV}}_{down} = (1.00, 1.00, 0.60, 0.83, 1.00)$ \\ [+0.25em] \hline \\ [-0.75em]
\multirow{16}{*}{Case 2}&   &  &   & $\hat{\mathbf{T}}_{down} = (3.45, 5.90, 4.11, 6.59, 9.96)$\\
& Est. 1  & 6    &0.54   & $\widehat{\mathbf{CV}}_{up}= (0.92, 0.82, 0.45, 0.77, 0.47)$\\
&     &    &        & $\widehat{\mathbf{CV}}_{down} = (0.34, 0.12, 0.58, 0.82, 0.11)$\\[+0.5em]
&   &   &   & $\hat{\mathbf{T}}_{down}= (5.45, 6.99, 5.71, 11.91, 9.94)$ \\
& Est. 2    & 8    & 0.43 & $\widehat{\mathbf{CV}}_{up}= (0.59, 0.46, 0.29, 0.12, 0.22)$\\
&     &     &       & $\widehat{\mathbf{CV}}_{down}= (0.11, 0.74, 0.67, 0.60, 0.51)$\\[+0.5em]
&  &  &   & $\hat{\mathbf{T}}_{down}= (9.38, 7.64, 17.40, 6.13, 9.45)$\\
& Est. 3    & 10   & 0.32 & $\widehat{\mathbf{CV}}_{up}= (0.47, 0.12, 0.27, 0.13, 0.26)$\\
&     &     &     & $\widehat{\mathbf{CV}}_{down}= (0.15, 0.46, 0.11, 0.37, 0.86)$\\[+0.5em]
&  &  &   & $\hat{\mathbf{T}}_{down}= (9.12, 5.60, 8.54, 8.59, 13.14)$ \\
& Est. 4    & 9   & 0.38  & $\widehat{\mathbf{CV}}_{up}= (0.21, 0.75, 0.23, 0.58, 0.11) $ \\
&     &    &        & $\widehat{\mathbf{CV}}_{down}= (0.35, 0.26, 0.67, 0.22, 0.38)$ \\ \hline \hline      
\end{tabular}
\end{table}

\begin{figure}[p]
    \centering    
    \subfloat[Objective function values resulting from different estimated parameters for $M=5$ cases under $N$ expansion]
      {\includegraphics[trim= 65 10 70 0,clip, width=0.9\linewidth]{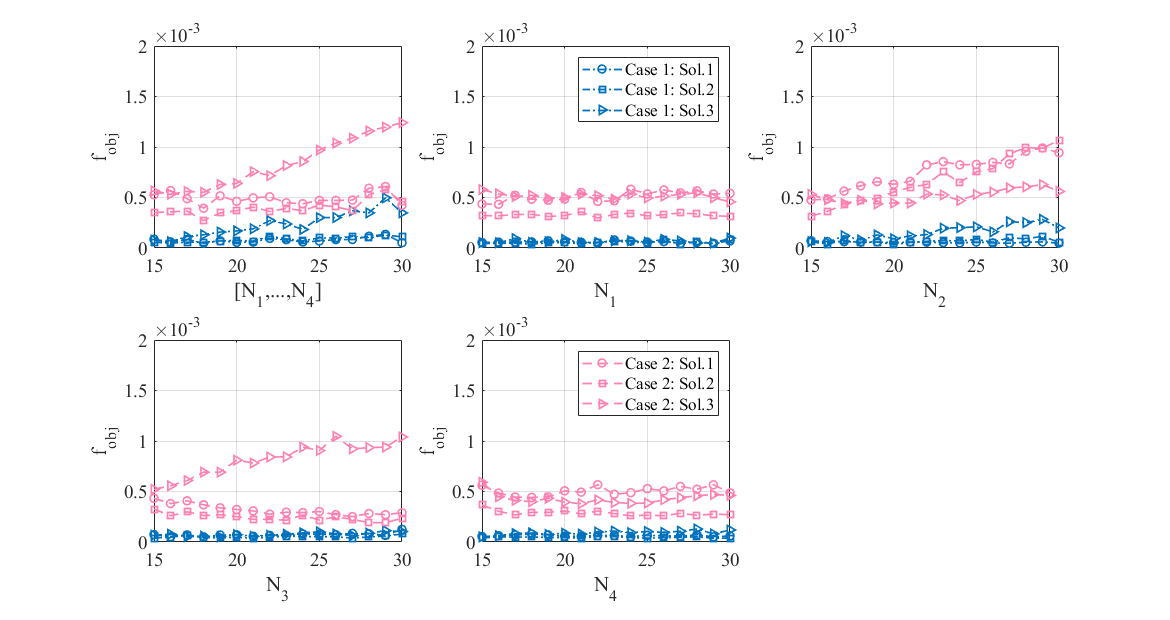}
         \label{fig_c2sa_obj_n}}\\   
         
     \subfloat[Objective function values resulting from different estimated parameters for $M=5$ cases under $T_{down}$ reduction]
         {\includegraphics[trim= 65 10 70 0,clip, width=0.9\linewidth]{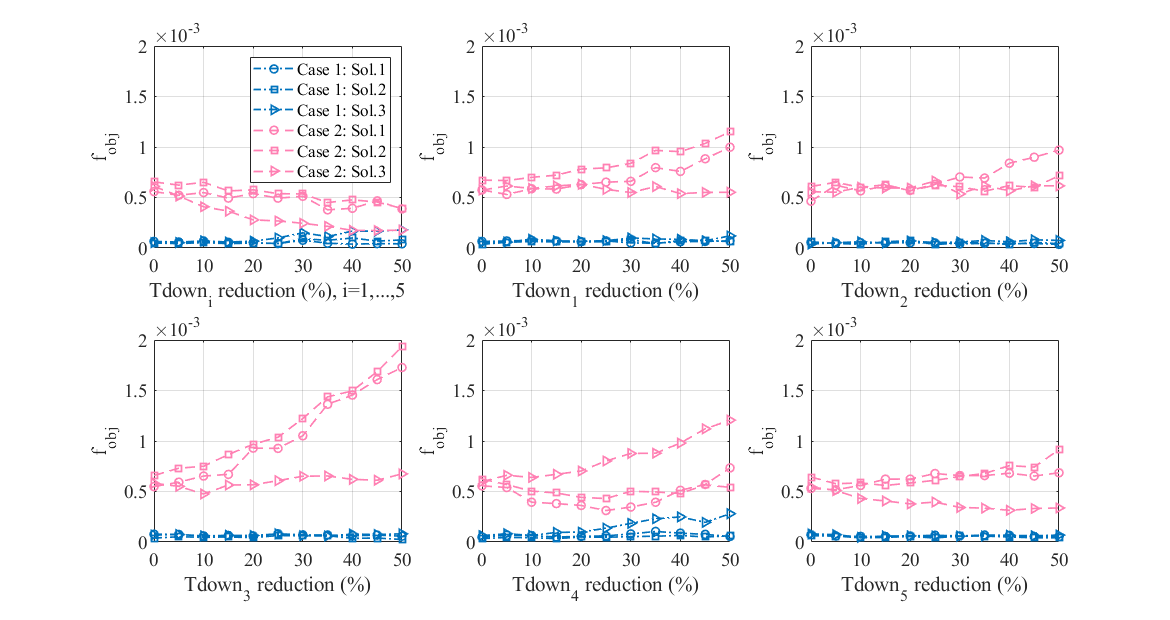}
         \label{fig_c2sa_obj_td}}\\   
\caption{Objective function values resulting from different estimated parameters for $M=5$ cases under improvement scenarios}
\label{fig_c2sa_obj}
\end{figure}

With these solutions of estimated machine parameters, 
we compare the estimated performance metrics resulting from them with the true ones, under the changes of both or individual buffer capacities from $N_j$ to $2N_j$, and also, the changes of all or individual $T_{down}$'s from the original values to 50\% reduction. 
For $M=5$, since the space limitation, we only show the plots of the objective function values in figures \ref{fig_c2sa_obj_n} and \ref{fig_c2sa_obj_td}. 
From all the figures, we can see that the estimated performance metrics are still very close to the true ones even though the buffer capacities are doubled or the $T_{down}$ is reduced to half.  Besides, we find that, for the cases of low CV systems, the performance metrics estimation errors resulting from the estimated parameters are usually higher than those of high CV systems.

More generally, we randomly select 1000 different cases for each $M=\{3,5\}$ and then for each case, we select 5 average estimation points and find the corresponding solution of estimated parameters. With these estimated parameters, we compute the estimated performance metrics under buffer expansion and $T_{down}$ reduction and compare them with the true ones. The errors are shown as Table\ref{tab_exp3_c1234_a}. 

\begin{table}[ht]
\centering
\small
\caption{Average estimation errors of performance metrics resulting from multiple estimated parameters under improvement scenarios}
\label{tab_exp3_c1234_a}
\begin{tabular}{llllll}
\hline \hline \\ [-0.85em]
  &Avg. $\epsilon$  & Double $N$'s  & Double $N_j$  & Half $T_{down}$'s  & Half $T_{down,j}$ \\ \hline \\[-0.85em]
\multirow{7}{*}{$M=3$} & $\epsilon_{PR}$  & 0.2529\%  & 0.2133\%  &0.1315\% & 0.3484\% \\
 &$\epsilon_{WIP_j}$ & 0.8083\% & 0.5556\% & 0.6829\% & 0.5547\%  \\
 &$\epsilon_{P_{0,j}}$  & 0.0043  & 0.0038 & 0.0038   & 0.0056   \\
 &$\epsilon_{P_{N,j}}$  &  0.0042 & 0.0034 & 0.0044   & 0.0058   \\
 &$\epsilon_{P_{Lk,j}}$  & 0.0058 & 0.0039 & 0.0056   & 0.0045     \\
 &$\epsilon_{B_{0,j}}$  &  0.3584\% & 0.2716\% & 0.2823\% & 0.3572\%  \\
 &$f_{obj}$  & $6.52\times 10^{-5}$  & $2.35\times 10^{-4}$ & $5.07\times 10^{-5}$ & $3.18\times 10^{-4}$ \\ [+0.25em] \hline \\ [-0.75em]
\multirow{7}{*}{$M=5$} & $\epsilon_{PR}$  & 0.3992\%  & 0.4093\%  & 0.2295\% & 0.5122\% \\
 &$\epsilon_{WIP_j}$ & 1.4747\%\% & 1.0562\% & 1.2062\% & 1.1488\%    \\
 &$\epsilon_{P_{0,j}}$  & 0.0083  & 0.0073 & 0.0080  & 0.0099     \\
 &$\epsilon_{P_{N,j}}$  &  0.0087 & 0.0074 & 0.0083   & 0.0101   \\
 &$\epsilon_{P_{Lk,j}}$  & 0.0094 & 0.0088 & 0.0103  & 0.0096      \\
 &$\epsilon_{B_{0,j}}$  &   0.6512\% & 0.5383\% & 0.4758\% & 0.6132\%  \\
 &$f_{obj}$  & $2.26\times 10^{-4}$ & $4.88\times 10^{-4}$ & $2.10\times 10^{-4}$  & $6.56\times 10^{-4}$  \\ \hline \hline
\end{tabular}
\end{table}

\subsubsection{Sensitivity analysis of model type } \label{type_sen_m}
As a special case of the gamma reliability model, the exponential model ($CV=1$) is more commonly used in production system research because the analytical methods are available. Given a set of performance metrics and With our proposed algorithm, we fix all $CV=1$, and then, we can obtain the optimal solution of estimated exponential parameters. In this subsection, the sensitivity of model type, (i.e., fit a gamma system as an exponential model) is investigated.

For $M=5$, we take 3 cases from Group 3 as examples, i.e.,\\
$e=(0.85, 0.85, 0.85, 0.85, 0.85)$, $N=(15, 15, 15, 15)$, and 
\begin{align*}
\text{Case 1:}\;& T_{down}=(9,7,12,10,12), (\bar{T}^*_{down}=10), \\
&CV = (0.99, 0.98, 0.76, 0.94, 0.81, 0.85, 0.89, 0.99, 0.85,  1.00), \bar{CV}^*_{avg}=0.9060\\
\text{Case 2:}\;& T_{down}=(6,8,7,10,9), (\bar{T}^*_{down}=8), \\
&CV = (0.80, 0.82, 0.87, 0.80, 0.96, 0.80, 0.80, 0.79, 0.80,    0.86), \bar{CV}^*_{avg}=0.8300\\
\text{Case 3:}\;& T_{down}=(8,6,10,9,12), (\bar{T}^*_{down}=9), \\
&CV = (0.52, 0.76, 0.59, 0.42, 0.72, 0.65, 0.91, 0.82, 0.60, 0.76), \bar{CV}_{avg}=0.6750\\
\text{Case 4:}\;& T_{down}=(8,12,10,6,9), (\bar{T}^*_{down}=9), \\
&CV = (0.46, 0.34, 0.36, 0.30, 0.39, 0.47, 0.43, 0.42, 0.41, 0.47), \bar{CV}_{avg}=0.4050.
\end{align*} 

We set all $CV$'s equal to 1, and then we can obtain the corresponding $T_{down}$ using our Algorithm (M-PSO). Note that, when all the CV's values are determined, the optimal solution of estimated $T_{down}$ is unique. The solutions of exponential parameters for five-machine cases are shown in Table \ref{tab_exp5_c1234}. For cases 1, 2, and 3, the valid solution of exponential parameters can be found, while no valid solution exists for case 4, the low CV systems. It is clear to see from figure \ref{fig_cv_m5}, for the low CV system, when $\bar{CV}_{avg}=1$, the corresponding $\bar{T}_{down}$ may be negative, which is infeasible. Thus, we cannot find any valid solutions for case 4.

\begin{table}[ht]
\centering
\small
\caption{Estimated exponential parameters of selected cases for $M=5$}
\label{tab_exp5_c1234}
\begin{tabular}{llcccc} 
\hline \hline \\[-0.95em]
                        & Estimated exponential parameters & NN-based $f_{obj}$   &Simulated $f_{obj}$     & $\epsilon_{PR}$ & $\epsilon_{WIP_j}$\\  [+0.05em] \hline \\ [-0.95em]
\multirow{2}{*}{case 1} & $\mathbf{\hat{e}} = (0.85, 0.85, 0.86, 0.85, 0.85)$   & \multirow{2}{*}{$1.89\times 10^{-5}$} & \multirow{2}{*}{$3.47\times 10^{-5}$} & \multirow{2}{*}{$0.64\%$} & \multirow{2}{*}{$0.38\%$}\\
                        & $\mathbf{\hat{T}_{down}}=(8.02,6.62,8.36,9.40,10.13)$ &                   &   & & \\ [+0.5em]
\multirow{2}{*}{case 2} & $\mathbf{\hat{e}} = (0.85, 0.86, 0.85, 0.86, 0.85)$   & \multirow{2}{*}{$9.53\times 10^{-6}$} & \multirow{2}{*}{$1.16\times 10^{-4}$} & \multirow{2}{*}{$0.81\%$} & \multirow{2}{*}{$0.64\%$}\\
                        & $\mathbf{\hat{T}_{down}}=(4.20,5.18,5.82,5.34,9.76)$  &                   & & & \\ [+0.5em]
\multirow{2}{*}{case 3} & $\mathbf{\hat{e}} = (0.86, 0.87, 0.87, 0.87, 0.86)$   & \multirow{2}{*}{$2.97\times 10^{-5}$} & \multirow{2}{*}{$1.62\times 10^{-4}$}& \multirow{2}{*}{$1.95\%$} & \multirow{2}{*}{$0.66\%$}\\
                        & $\mathbf{\hat{T}_{down}}=(3.93,3.83,5.20,3.47,8.61)$  &                   & & & \\ [+0.5em]
\multirow{2}{*}{case 4*} & $\mathbf{\hat{e}} = (0.86, 0.88, 0.88, 0.88, 0.87)$  & \multirow{2}{*}{$4.88\times 10^{-4}$} & \multirow{2}{*}{$1.42\times 10^{-3}$}& \multirow{2}{*}{$3.17\%$} & \multirow{2}{*}{$4.23\%$}\\
                        & $\mathbf{\hat{T}_{down}}=(3.52,2.44,2.00,2.00,2.01)$  &     &  & &
                        \\ \hline  \hline
\multicolumn{4}{l}{\footnotesize *The optimal solution of case 4 is not valid.} 
\end{tabular}
\end{table}

\begin{figure}[p]
    \centering    
    \subfloat[Simulation-based objective function values resulting from estimated exponential parameters for $M=5$ cases under $N$ expansion]
      {\includegraphics[trim= 70 10 60 0,clip, width=0.8\linewidth]{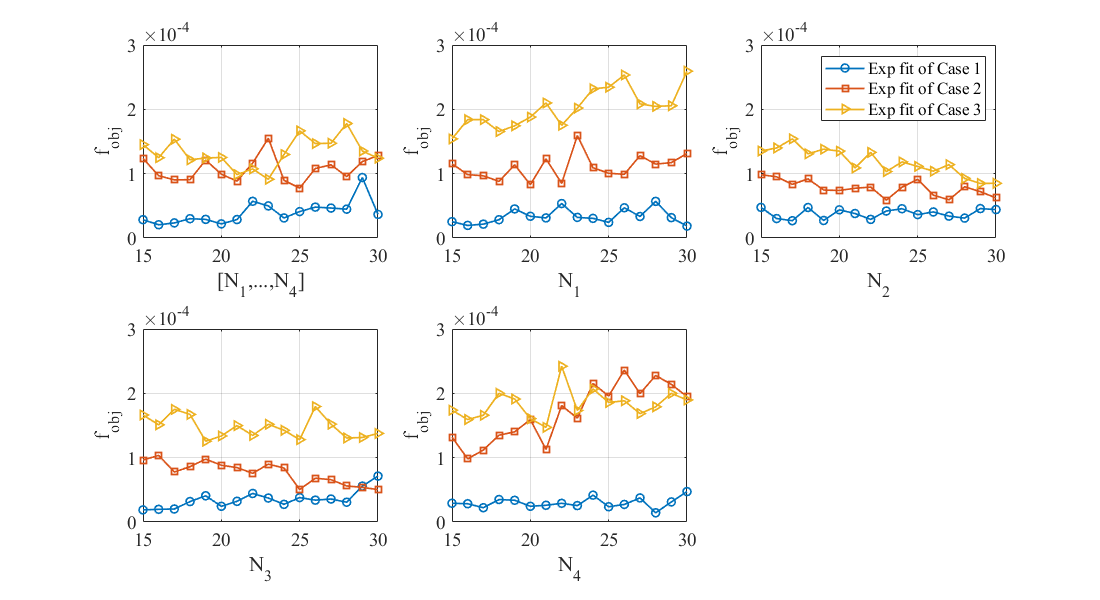}
         \label{fig1_g5obj_n}}\\   
         
     \subfloat[Simulation-based objective function values resulting from estimated exponential parameters for $M=5$ cases under $T_{down}$ reduction]
         {\includegraphics[trim= 70 10 60 0,clip, width=0.8\linewidth]{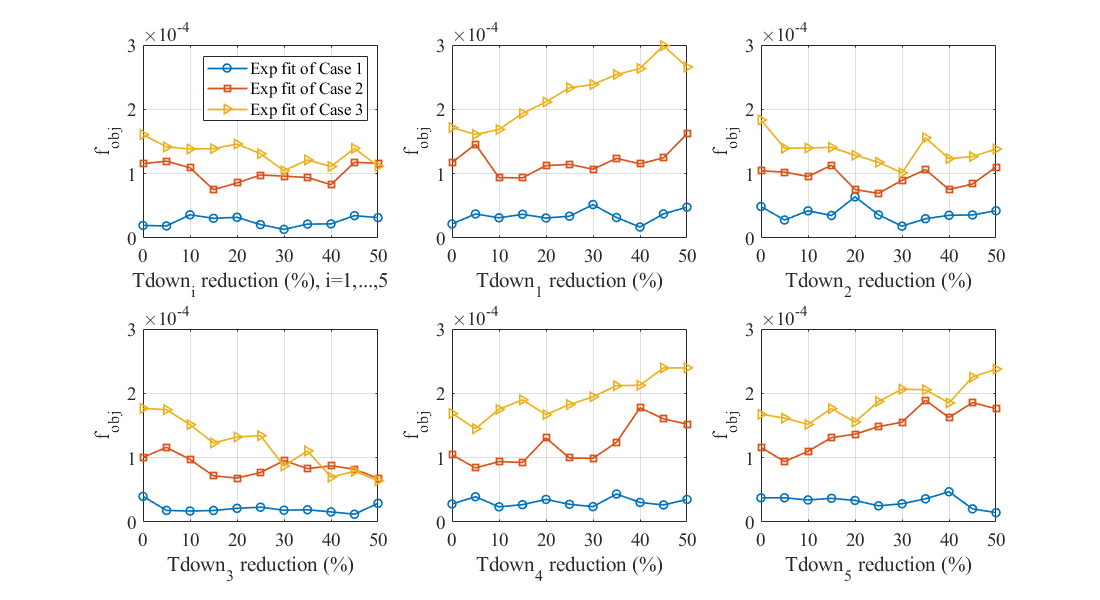}
         \label{fig_g5obj_td}}\\   
\caption{Objective function values resulting from estimated exponential parameters for $M=5$ cases under improvement scenarios}
\label{fig_exp2sa5_obj}
\end{figure}

Similar to subsection \ref{par_sen_p}, we investigate the estimated performance metrics under buffer expansion and downtime reduction. For case 4, neither NN-based nor simulated $f_{obj}$ is below the threshold of valid estimation ($10^{-4}$), and we obtain the errors of corresponding performance metrics are much higher than those of the other 3 cases. Moreover, when the buffer capacities are doubled or the downtimes are reduced to half, the maximum of $\epsilon_{PR}$ and $\epsilon_{WIP_j}$ increase to about 7\%. Case 4 is not recommended to fit as an exponential model. Thus, we exclude case 4 and only plot the objective function values under those improvement scenarios resulting from the exponential fits of cases 1, 2, and 3 as shown in Figure \ref{fig_exp2sa5_obj}.

According to Figure \ref{fig_exp2sa5_obj}, there is no significant increase in $f_{obj}$ when buffer expansion or downtime reduction is carried out for these systems. 
That means, for cases 1, 2, and 3, the exponential models obtained from our proposed method can also fit the systems very well, although the true distribution of their up- and downtimes is not exponential. Besides, since the system's overall CV of case 1 is the highest and close to 1, the corresponding exponential fit has much lower errors of estimated performance metrics than the other two cases. Furthermore, the exponential fit for case 3 has the highest estimation errors, because the overall CV of this system is the lowest.

\begin{figure}[p]
\centering    
    \subfloat[Average $f_{obj}$ resulting from the estimated exponential parameters with different true $\bar{CV}_{avg}$ and $\bar{T}_{down}$ for $M=3$]
    {\includegraphics[trim= 30 0 30 0, clip, width=0.65\linewidth]{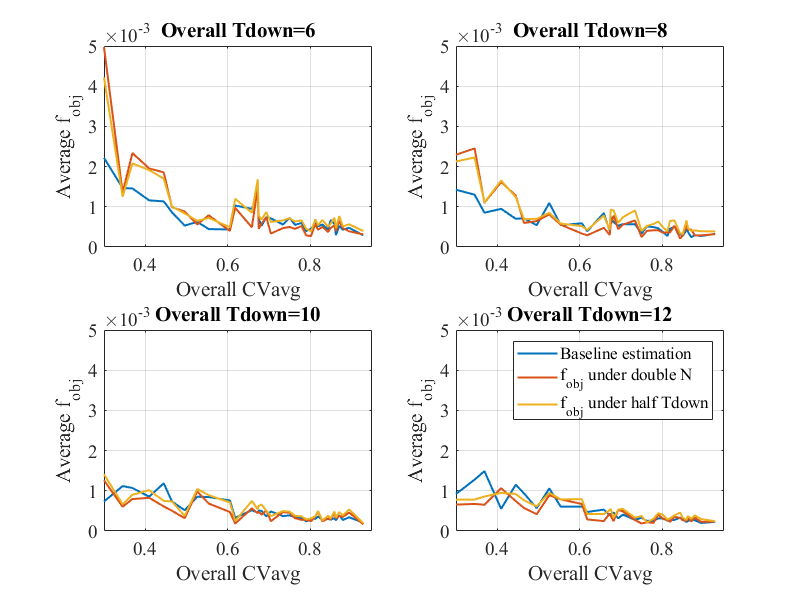}
     \label{fig1_g3_expF}} \\      
    \subfloat[Average $f_{obj}$ resulting from the estimated exponential parameters with different true $\bar{CV}_{avg}$ and $\bar{T}_{down}$ for $M=5$]
    {\includegraphics[trim= 30 0 30 0, clip, width=0.65\linewidth]{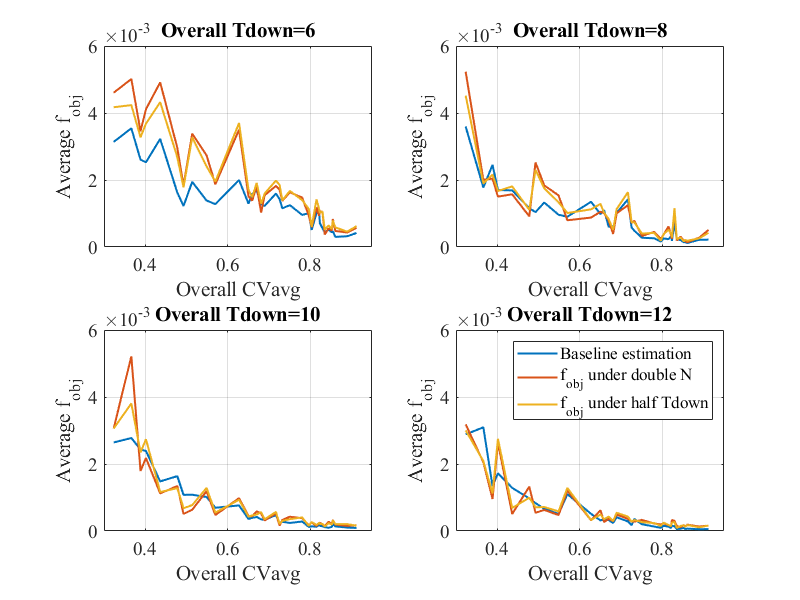}
     \label{fig1_g5_expF}}
     
\caption{Average $f_{obj}$ resulting from the estimated exponential parameters with different true $\bar{CV}_{avg}$ and $\bar{T}_{down}$}
\label{fig1_expF}
\end{figure}

More general, we set $\bar{T}_{down} = \{6, 8, 10, 12\}$ and randomly select 35 different values of 
$\bar{CV}_{avg}\in(0.25,0.95)$. For each combination of ($\bar{T}_{down}, \bar{CV}_{avg}$), we assign 10 different sets of the values of $T_{down,i}$'s, $CV_{up,i}$'s and $CV_{down,i}$'s which must match their average. So, we obtain $4\times35\times10= 1400$ cases, and for each case, machine efficiencies and buffer capacities are all randomly selected. Given the performance metrics data, we fit them as exponential models in the same way we introduced below. Then, we estimate the performance metrics with these exponential parameters on baseline, and under the improvement scenarios of double $N$ and half $T_{down}$ for all machines. In figure \ref{fig1_expF}, sorting by the true $\bar{CV}_{avg}$, We plot the average values of the objective function, which reflect the errors of estimated performance metrics compared with the true ones.

We can see that the errors of estimated performance metrics resulting from the exponential fits decrease with $\bar{CV}_{avg}$. On the high level of $\bar{CV}_{avg}$, these exponential fits lead to very similar and very low errors under all $\bar{T}_{down}$. However, for the low level of $\bar{CV}_{avg}$ system, the errors from the exponential fits have a significant increase with the decrease of $\bar{T}_{down}$, e.g.,  $\bar{T}_{down}=6$.

\section{Conclusions}
\label{sec_con}
In this paper, we apply the parts flow performance metrics-based modeling approach to identify the mathematical models for 
synchronous non-exponential production serial lines. Specifically, we estimate the non-exponential model parameters by matching the 
system performance metrics (e.g., production rate, work-in-process, the probability distribution of buffer occupancy, etc.) which can be derived from the parts flow data collected from the buffer. Since no close-formed analytical expression of performance metrics in terms of system parameters is available, we first build neural network surrogate models to perform the analytical expressions-based calculation of system performance metrics instead of using time-consuming simulations. 
Then, a constrained optimization problem is formulated to find the optimized solutions of estimated machine parameters that minimize the mean square error of the resulting estimated performance metrics to the true ones.  
To solve this optimization problem efficiently, a multi-start particle swarm optimization (M-PSO) algorithm with tightened constraints is designed. According to the numerical experiments, using the proposed M-PSO algorithm, we can find multiple non-unique solutions of estimated machine parameters that lead to practically the same performance metrics compared to those observed under true machine parameters. 
Furthermore, based on the multiple estimations, the negative linear patterns of overall downtime vs. overall CV leading to the same system performance metrics are observed and then analyzed.   
Finally, the validity and robustness of the models are investigated through extensive numerical experiments of model sensitivity analysis. 

In future work, this parts flow performance metrics-based modeling approach will be extended to more complex production system models, such as asynchronous production line models, assembly system models, etc. Furthermore, we will test the theoretical results in a simulated lab environment and continue our efforts in applying this approach in real manufacturing systems.

\section*{Acknowledgement}
This work is supported in part by the U.S. National Science Foundation (NSF), under Grant number FM-2134367.

\section*{Disclosure statement}
Dr. Liang Zhang has financial interests and/or other relationships with Smart Production Systems LLC, Ann Arbor, MI, USA.

\bibliographystyle{unsrt}  
\bibliography{references}

\vspace{10pt}
\begin{Large}
\textbf{Appendix}
\end{Large}

\appendix
\section{Distribution of estimated machine parameters: illustration examples of three-machine cases by groups}
\subsection{Group 1} See the main text.
\subsection{Group 2} All machines have different $T_{down}$'s and $CV$'s, but $\bar{T}_{down}$ or $\bar{CV}_{avg}$ are set as $\bar{T}_{down} = \{6,8,10,12\}$, $\bar{CV}_{avg}$=\{0.5, 0.75\}. Specifically, for $M=3$, $e=[0.85, 0.85, 0.85]$, $N=[15, 15]$, and 
\begin{align*}
T_{down}1 =& \{(7,6,5), (5,6,7)\},\;(\bar{T}_{down}=6)\\ 
T_{down}2 =& \{(6,8,10), (11,7,6)\},\;(\bar{T}_{down}=8)\\ 
T_{down}3 =& \{(10,9,11), (8,10,12)\},\;(\bar{T}_{down}=10)\\ 
T_{down}4 =& \{(12,10,14), (11,13,12)\},\;(\bar{T}_{down}=12)\\
CV1 = &\{(0.6, 0.7, 0.45, 0.35, 0.4, 0.5), (0.45, 0.3, 0.75, 0.5, 0.35, 0.65)\}, (\bar{CV}_{avg}=0.5)\\
CV2 = &\{(0.7, 0.85, 0.8, 0.95, 0.65, 0.55), (0.55, 0.7, 0.65, 0.75, 1, 0.85)\}, (\bar{CV}_{avg}=0.75)
\end{align*}

\begin{figure}[ht]
\centering
{{\includegraphics[trim= 0 0 15 0,clip, scale=0.475]{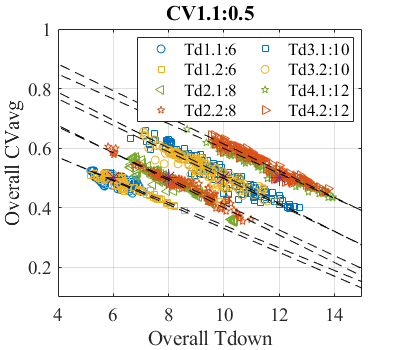}}
{\includegraphics[trim= 0 0 15 0,clip, scale=0.475]{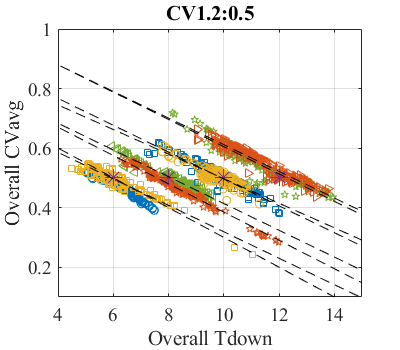}}\\
{\includegraphics[trim= 0 0 15 0,clip, scale=0.475]{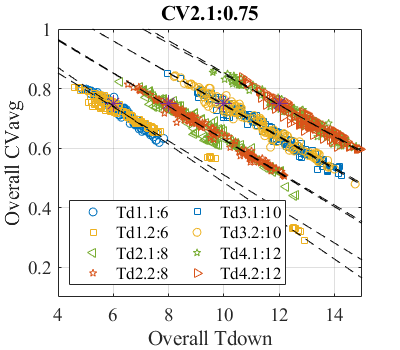}}
{\includegraphics[trim= 0 0 15 0,clip, scale=0.475]{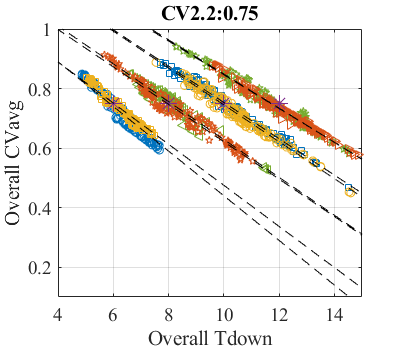}}}
\caption{Overall $T_{down}$ vs. $CV_{avg}$ for estimated parameters of Group 2 for $M=3$}
\label{fig_avg_t_cv_m3}
\end{figure}

\subsection{Group 3}
All machines have different $T_{down}$'s and $CV$'s are grouped by low, medium, and high levels. For $M=3$, we set $e=[0.85, 0.85, 0.85]$, $N=[15, 15]$, and
\begin{align*}
&T_{down}1=(6,12,6), \bar{T}_{down}=8\\
&T_{down}2=(6,9,12), \bar{T}_{down}=9\\
&T_{down}3=(12,9,6), \bar{T}_{down}=9\\
&T_{down}4=(12,6,12), \bar{T}_{down}=10
\end{align*}
Low CV: 
\begin{align*}
    &CV1 = (0.41,0.32,0.33,0.35,0.47,0.35), \bar{CV}_{avg}=0.3717\\
    &CV2 = (0.32,0.40,0.50,0.37,0.42,0.34), \bar{CV}_{avg}=0.3917\\
    &CV3 = (0.45,0.35,0.40,0.44,0.48,0.50), \bar{CV}_{avg}=0.4367
\end{align*}
Medium CV: 
\begin{align*}
    &CV1 = (0.82, 0.8, 0.3, 0.35, 0.7, 0.66), \bar{CV}_{avg}=0.6050\\
    &CV2 = (0.8, 0.35, 0.7, 0.72, 0.83, 0.37), \bar{CV}_{avg}=0.6283\\
    &CV3 = (0.73, 0.32, 0.84, 0.80, 0.91, 0.35), \bar{CV}_{avg}=0.6583
\end{align*}
High CV: 
\begin{align*}
    &CV1 = (0.87,0.77,0.81,0.87,0.88,0.82), \bar{CV}_{avg}=0.8367\\
    &CV2 = (0.96,0.84,0.81,0.90,0.93,0.82), \bar{CV}_{avg}=0.8767\\
    &CV3 = (0.97,0.90,0.98,0.79,0.96,0.84), \bar{CV}_{avg}=0.9067
\end{align*}

\begin{figure}[ht]
     \centering
     \subfloat[Estimated parameters for low CV cases ]
         {\includegraphics[trim= 70 0 70 0,clip, width=0.9\linewidth]{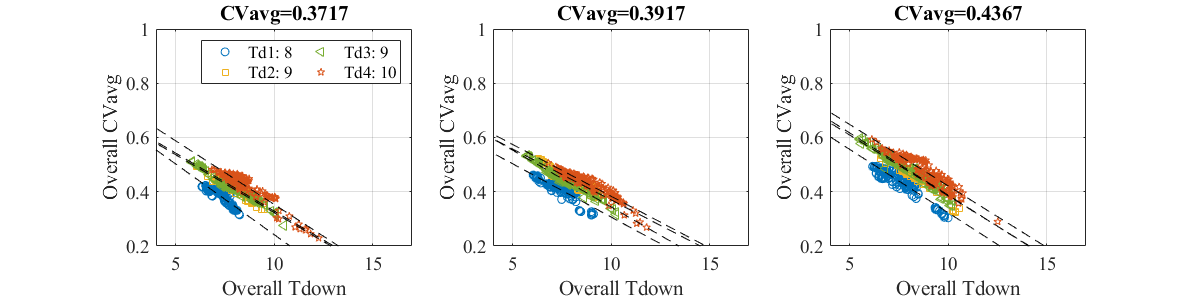}
         \label{fig1_tcva_L1}}\\
         
    \subfloat[Estimated parameters for medium CV cases]
         {\includegraphics[trim= 70 0 70 0,clip, width=0.9\linewidth]{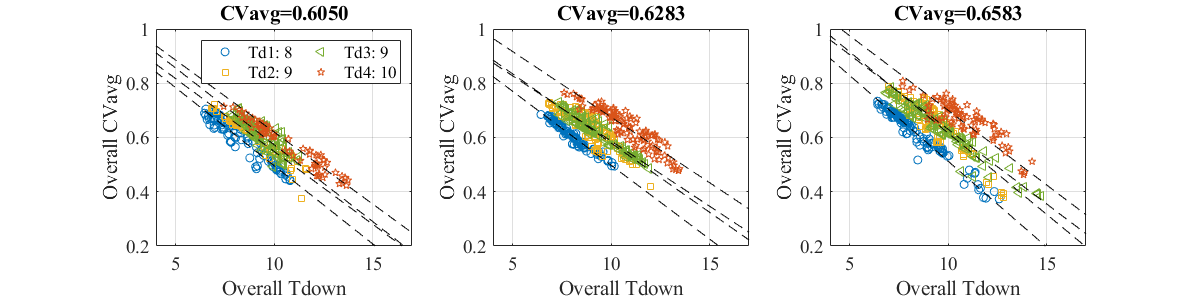}
         \label{fig1_tcva_M1}}\\ 

     \subfloat[Estimated parameters for high CV cases]
         {\includegraphics[trim= 70 0 70 0,clip, width=0.9\linewidth]{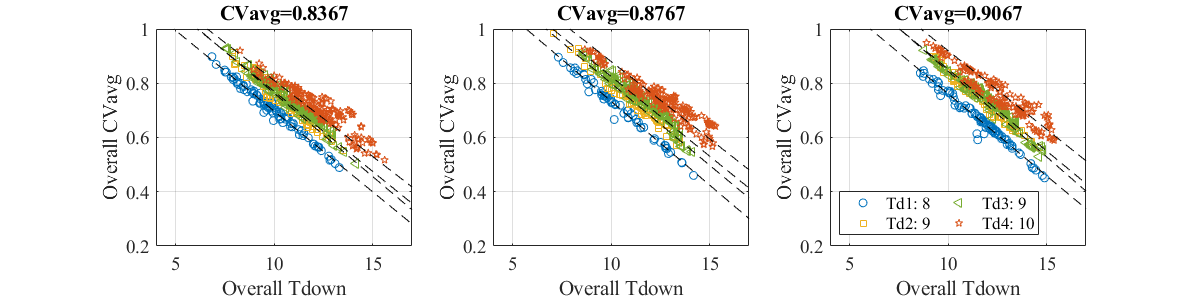}
         \label{fig1_tcva_H1}}\\         
    \caption{Overall $T_{down}$ vs. $CV_{avg}$ for estimated parameters of $M=3$ cases in Group 3}
        \label{fig_cv_m3}
\end{figure}

\subsection{Group 4} See the main text.

\section{Sensitivity analysis of model parameters: illustration examples of three-machine cases}

According to numerical facts 2 and 3, we first find the fitted linear function for a case. Then, we randomly select several average estimation points and search for valid solutions with the average estimation constraints. 
For instance, we take two 3-machine cases from Group 3 as examples, i.e.,\\
$e=(0.85, 0.85, 0.85)$, $N=(15, 15)$, $T_{down}=(6,9,12)$ ($\bar{T}^*_{avg}=9$)\\
Case 1: $CV = (0.97,0.90,0.98,0.79,0.96,0.84)$, $\bar{CV}^*_{avg}=0.9067$\\
Case 2: 
$CV = (0.41,0.32,0.33,0.35,0.47,0.35)$, $\bar{CV}^*_{avg}=0.3717$\\
The fitted linear functions of these 2 cases are\\
Case 1: $y=-0.0605x + 1.4522$\\
Case 2: $y=-0.0408x + 0.7392$\\
We select several average estimation points from the fitted linear function and obtain the following valid solutions in Table \ref{tab_est3_c12}. Since all the estimated machine efficiencies obtained for these cases are very close to the true ones (i.e., all $\hat{e}\approx 0.85$), they are not shown in Table \ref{tab_est3_c12}.

\begin{table}[ht]
\centering
\small
\caption{Selected estimations of $M=3$ cases 1 and 2}
\label{tab_est3_c12}
\begin{tabular}{llccl}
\hline\hline \\ [-0.75em]
&  & \multicolumn{1}{c}{$\bar{T}_{down}$}  & \multicolumn{1}{c} {$\bar{CV}_{avg}$} & \multicolumn{1}{c}{Solutions}\\   \hline \\[-0.75em]
\multirow{15}{*}{Case 1}&  &  &   & $\hat{\mathbf{T}}_{down} = (5.17,8.11,10.72)$ \\
&Est. 1 & 8 & 0.97  & $\widehat{\mathbf{CV}}_{up} = (1.00, 1.00, 1.00)$ \\
&     &    &       & $\widehat{\mathbf{CV}}_{down} = (0.90, 1.00, 0.91)$ \\  [+0.5em]
&     &    &       & $\hat{\mathbf{T}}_{down} = (8.51, 7.71, 13.78)$  \\
&Est. 2 & 10 & 0.85 & $\widehat{\mathbf{CV}}_{up}= (0.58, 0.98, 0.85)$  \\
&     &     &       & $\widehat{\mathbf{CV}}_{down}= (0.93, 0.85, 0.90)$ \\ [+0.5em]
&     &     &       & $\hat{\mathbf{T}}_{down} = (7.24, 13.52, 15.24)$  \\
&Est. 3 & 12 & 0.73 & $\widehat{\mathbf{CV}}_{up} = (0.67, 0.67, 0.59)$  \\
&     &     &     & $\widehat{\mathbf{CV}}_{down} = (0.73, 0.93, 0.77)$ \\ [+0.5em]
&     &     &     & $\hat{\mathbf{T}}_{down} = (6.02, 8.84, 12.14)$    \\
&Est. 4 & 9 & 0.91 & $\widehat{\mathbf{CV}}_{up} = (0.87, 0.99, 0.90)$   \\
&     &    &        & $\widehat{\mathbf{CV}}_{down} = (0.85, 0.99, 0.84)$ \\ [+0.25em] \hline \\ [-0.75em]
\multirow{15}{*}{Case 2}&  &  &  & $\hat{\mathbf{T}}_{down} = (5.66, 4.49, 7.84)$ \\
&Est. 1    & 6  & 0.49 & $\widehat{\mathbf{CV}}_{up}= (0.38, 0.91, 0.67)$\\
&     &    &        & $\widehat{\mathbf{CV}}_{down} = (0.26, 0.46, 0.29)$\\ [+0.5em]
&     &    &       & $\hat{\mathbf{T}}_{down}= (5.64, 7.54, 10.82)$ \\
&Est. 2    & 8& 0.41 & $\widehat{\mathbf{CV}}_{up}= (0.31, 0.47, 0.22)$\\
&     &    &       & $\widehat{\mathbf{CV}}_{down}= (0.43, 0.59, 0.47)$\\ [+0.5em]
&     &    &      & $\hat{\mathbf{T}}_{down}= (5.96, 11.18, 12.86)$\\
&Est. 3  & 10& 0.33& $\widehat{\mathbf{CV}}_{up}= (0.47, 0.11, 0.24)$\\
&     &    &     & $\widehat{\mathbf{CV}}_{down}= (0.43, 0.40, 0.34)$\\ [+0.5em]
&     &    &     & $\hat{\mathbf{T}}_{down}= (5.72, 9.19, 12.09)$ \\
&Est. 4  & 9& 0.37  & $\widehat{\mathbf{CV}}_{up}= (0.42, 0.17, 0.22) $ \\
&     &    &        & $\widehat{\mathbf{CV}}_{down}= (0.43, 0.55, 0.45)$ \\ \hline \hline      
\end{tabular}
\end{table}

\begin{figure}[ht]
\centering
\includegraphics[trim= 70 0 70 0, clip, width=0.95\linewidth]{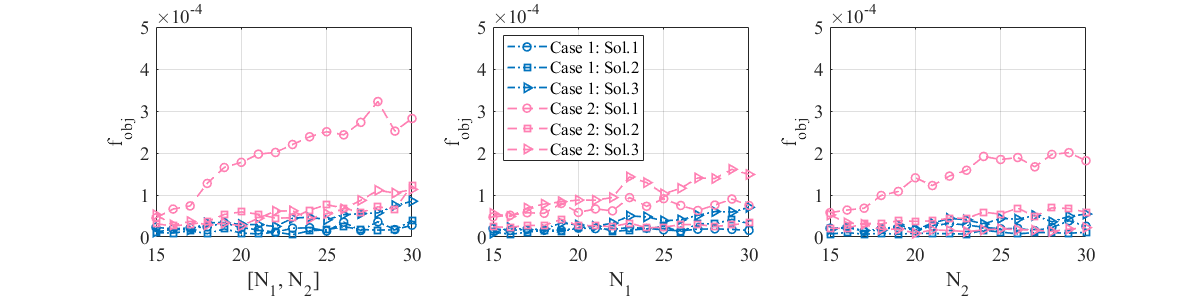}
\caption{Objective function values resulting from different estimated parameters for $M=3$ cases under $N$ expansion}
\label{fig_c1sa_obj_n}
\end{figure}

\begin{figure}[ht]
     \centering
     \subfloat[Estimated $PR$ of different estimated parameters under $N$ expansion]
         {\includegraphics[trim= 70 0 70 0,clip, width=0.85\linewidth]{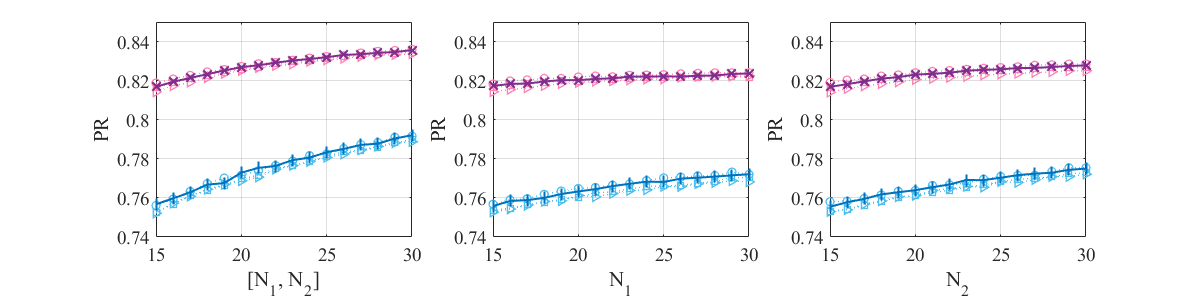}
         \label{fig1_c1sa_pr}}\\
     \subfloat[Estimated $WIP_1$ of different estimated parameters under $N$ expansion]
         {\includegraphics[trim= 70 0 70 0,clip, width=0.85\linewidth]{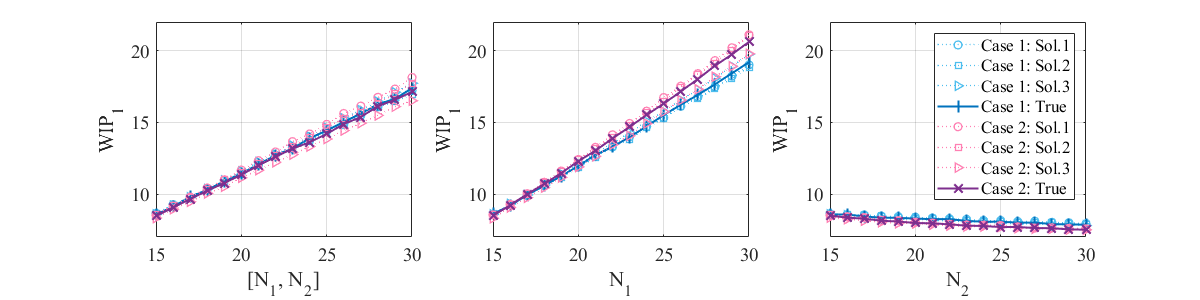}
         \label{fig1_c1sa_wip1}}\\
     \subfloat[Estimated $WIP_2$ of different estimated parameters under $N$ expansion]
         {\includegraphics[trim= 70 0 70 0,clip, width=0.85\linewidth]{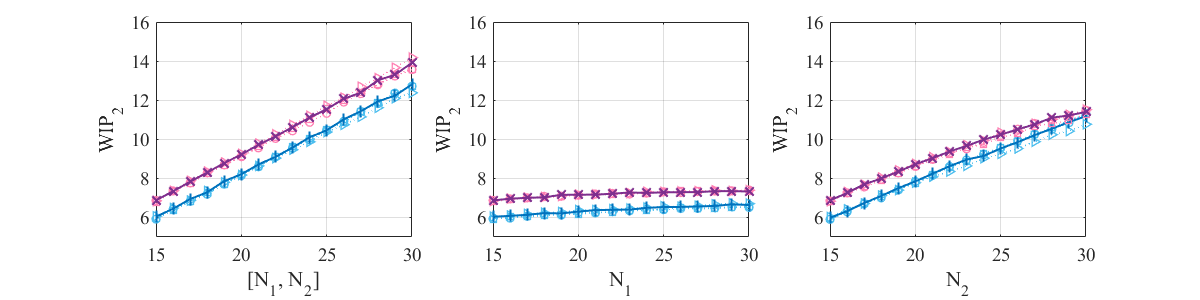}
         \label{fig1_c1sa_wip2}}\\ 
             
    \caption{Performance metrics of different estimated parameters for $M=3$ cases under $N$ expansion}
        \label{fig_m3c1sa_n}
\end{figure}

\begin{figure}[ht]
     \centering
     \subfloat[Estimated $PR$ of different estimated parameters under $T_{down}$ reduction]
         {\includegraphics[trim= 95 0 90 0,clip, width=1\linewidth]{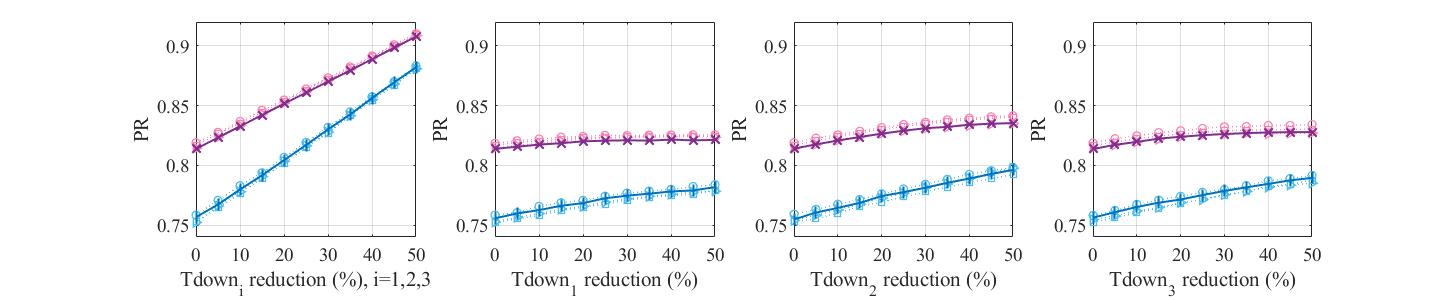}
         \label{fig1_c1sa_pr_td}}\\
     \subfloat[Estimated $WIP_1$ of different estimated parameters under $T_{down}$ reduction]
         {\includegraphics[trim= 95 0 90 0,clip, width=1\linewidth]{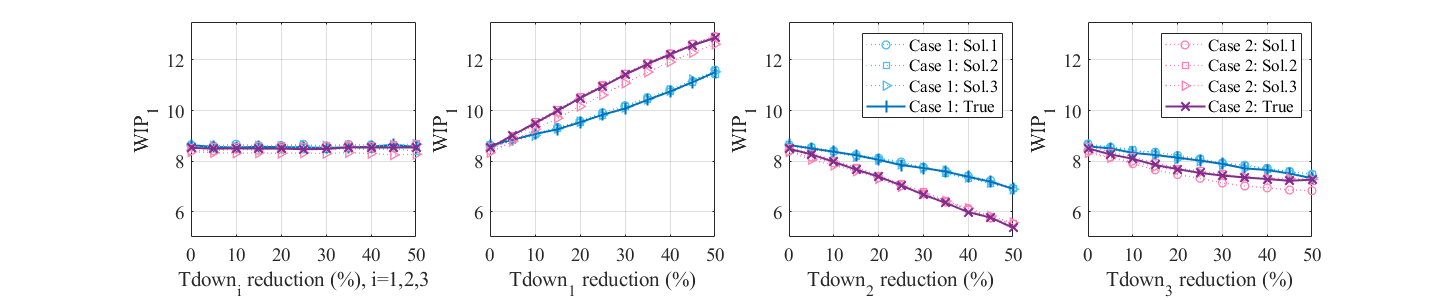}
         \label{fig1_c1sa_wip1_td}}\\
     \subfloat[Estimated $WIP_2$ of different estimated parameters under $T_{down}$ reduction]
         {\includegraphics[trim= 95 0 90 0,clip, width=1\linewidth]{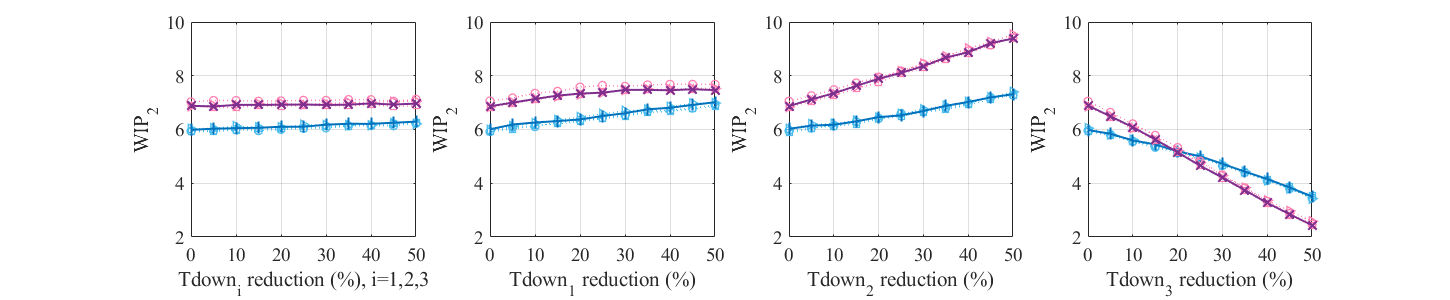}
         \label{fig1_c1sa_wip2_td}}\\     
\caption{Performance metrics of different estimated parameters for $M=3$ cases under $T_{down}$ reduction}
\label{fig_m3c1sa_td}
\end{figure}

\begin{figure}[ht]
\centering
\includegraphics[trim= 100 0 90 0, clip, width=1\linewidth]{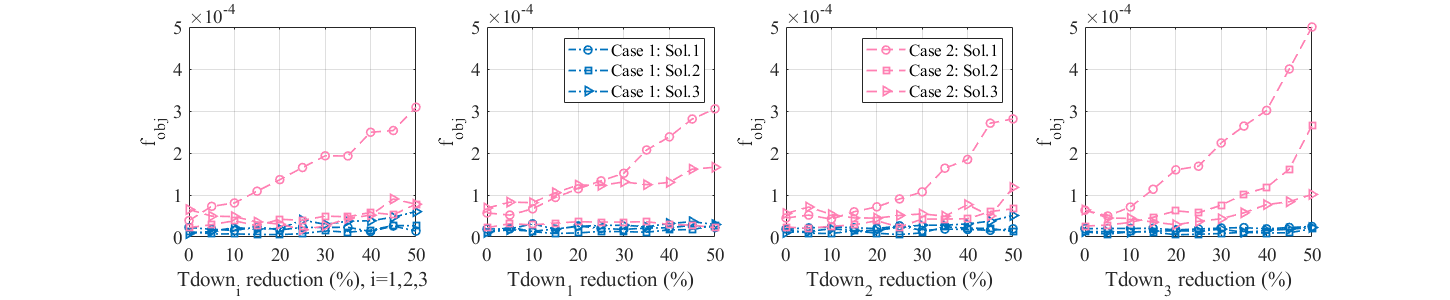}
\caption{Objective function values resulting from different estimated parameters for $M=3$ cases under $T_{down}$ reduction}
\label{fig_c1sa_obj_td}
\end{figure}

With these solutions of estimated machine parameters, 
we compare the estimated performance metrics resulting from them with the true ones, under the changes of both or individual buffer capacities from $N_i$ to $2N_i$, and also, the changes of all or individual $T_{down}$'s from the original values to 50\% reduction.    
For $M=3$, figures \ref{fig_m3c1sa_n} and \ref{fig_m3c1sa_td} show the estimated performance metrics, $PR$ and $WIP_i$'s, resulting from those estimated parameters in Table \ref{tab_est3_c12}. To evaluate the total errors of all estimated performance metrics. we also plot the objective function value (the mean square error) under buffer expansion and downtime reduction in figures \ref{fig_c1sa_obj_n} and \ref{fig_c1sa_obj_td}. 
we can see that the estimated performance metrics are still very close to the true ones even though the buffer capacities are doubled or the $T_{down}$ is reduced to half.

\section{Sensitivity analysis of model type: illustration examples of three-machine cases}

For $M=3$, we take 3 cases from Group 3 as examples, 
i.e., \\ $e=(0.85, 0.85, 0.85)$, $N=(15, 15)$, and 
\begin{align*}
\text{Case 1:}\;& T_{down}=(12,6,12), (\bar{T}^*_{down}=10), \\
&CV = (0.97, 0.90, 0.98, 0.79, 0.96, 0.84), \bar{CV}^*_{avg}=0.9067\\
\text{Case 2:}\;& T_{down}=(6,12,6), (\bar{T}^*_{down}=8), \\
&CV = (0.87, 0.77, 0.81, 0.87, 0.88, 0.82), \bar{CV}^*_{avg}=0.8367\\
\text{Case 3:}\;& T_{down}=(6,9,12), (\bar{T}^*_{down}=9), \\
&CV = (0.73, 0.32, 0.84, 0.80, 0.91, 0.35), \bar{CV}_{avg}=0.6583\\
\text{Case 4:}\;& T_{down}=(12,9,6), (\bar{T}^*_{down}=9), \\
&CV = (0.32, 0.40, 0.50, 0.37, 0.42, 0.34), \bar{CV}_{avg}=0.3917.
\end{align*}

We set all $CV$'s equal to 1, and then we can obtain the corresponding $T_{down}$ using our Algorithm (M-PSO). Note that, when all the CV's values are determined, the optimal solution of estimated $T_{down}$ is unique. The solutions of exponential parameters for three-machine cases are shown in Tables \ref{tab_exp3_c1234}. For cases 1, 2, and 3, the valid solution of exponential parameters can be found, while no valid solution exists for case 4, the low CV systems.

\begin{table}[ht]
\centering
\small
\caption{Estimated exponential parameters of selected cases for $M=3$}
\label{tab_exp3_c1234}
\begin{tabular}{llcc} 
\hline \hline \\[-0.95em]
                        & Estimated exponential parameters & NN-based $f_{obj}$  & Simulated $f_{obj}$     \\  [+0.05em] \hline \\ [-0.95em]
\multirow{2}{*}{case 1} & $\mathbf{\hat{e}} = (0.85, 0.85, 0.85)$   & \multirow{2}{*}{$1.51\times 10^{-5}$} & \multirow{2}{*}{$3.13\times 10^{-5}$}\\
                        & $\mathbf{\hat{T}_{down}}=(11.01, 4.57, 10.85)$ &                   &\\ [+0.5em]
\multirow{2}{*}{case 2} & $\mathbf{\hat{e}} = (0.85, 0.86, 0.85)$   & \multirow{2}{*}{$2.49\times 10^{-5}$} & \multirow{2}{*}{$4.58\times 10^{-5}$}\\
                        & $\mathbf{\hat{T}_{down}}=(5.15, 7.88, 4.83)$  &                   &\\ [+0.5em]
\multirow{2}{*}{case 3} & $\mathbf{\hat{e}} = (0.86, 0.87, 0.85)$   & \multirow{2}{*}{$9.98\times 10^{-5}$} & \multirow{2}{*}{$1.89\times 10^{-4}$}\\
                        & $\mathbf{\hat{T}_{down}}=(3.51, 2.92, 9.81)$  &                   &\\ [+0.5em]
\multirow{2}{*}{case 4*} & $\mathbf{\hat{e}} = (0.87, 0.88, 0.87)$   & \multirow{2}{*}{$1.03\times 10^{-3}$} & \multirow{2}{*}{$1.81\times 10^{-3}$}\\
                        & $\mathbf{\hat{T}_{down}}=(2.24, 2.41, 2.00)$  &     &
                        \\ \hline  \hline
\multicolumn{4}{l}{\footnotesize *The optimal solution of case 4 is not valid.} 
\end{tabular}
\end{table}

\begin{figure}[ht]
\centering
\includegraphics[trim= 70 0 70 0,clip, width=0.85\linewidth]{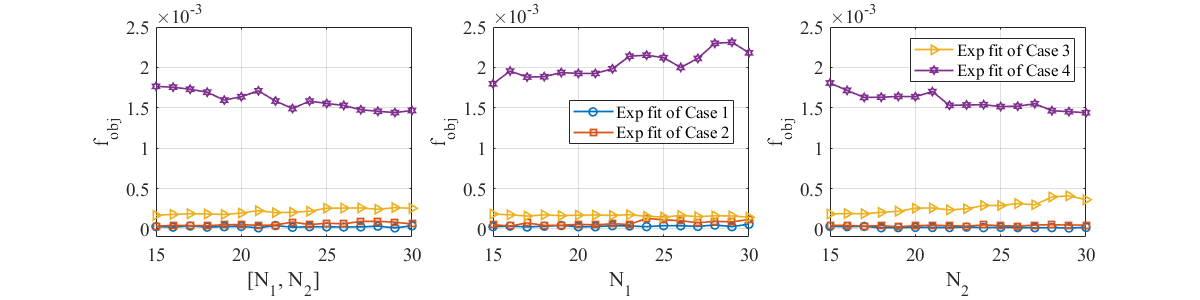}
\caption{$f_{obj}$ resulting from the estimated exponential parameters for $M=3$ under $N$ expansion}
\label{fig1_g3obj_n}
\end{figure}

\begin{figure}[ht]
     \centering
     \subfloat[Estimation errors of $PR$ resulting from estimated exponential parameters under $N$ expansion]
         {\includegraphics[trim= 70 0 70 0,clip, width=0.85\linewidth]{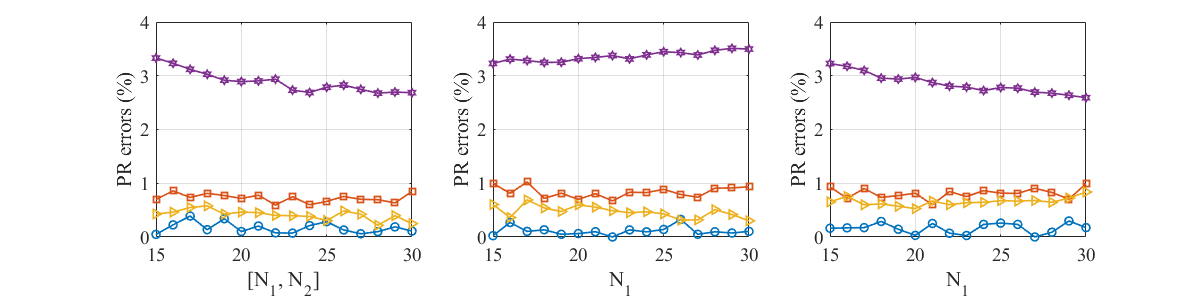}
         \label{fig1_exp1sa_pr}}\\
    \subfloat[Estimation errors of $WIP_{avg}$ resulting from estimated exponential parameters under $N$ expansion]
         {\includegraphics[trim= 70 0 70 0,clip, width=0.85\linewidth]{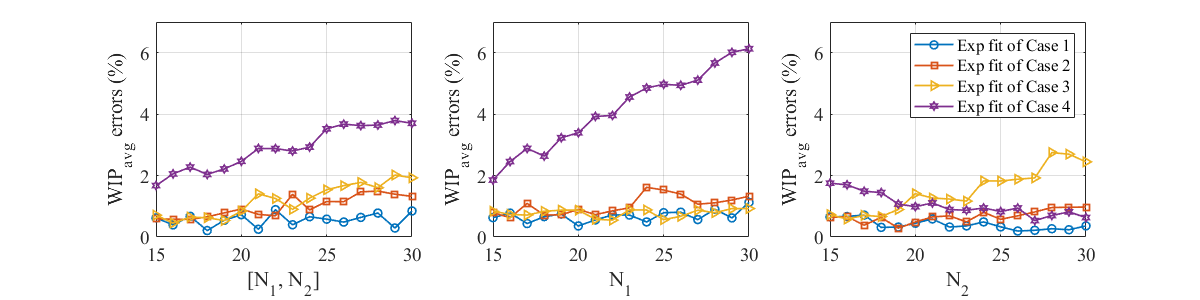}
         \label{fig1_exp1sa_wipa}}\\     
    \caption{Performance metrics errors resulting from the estimated exponential parameters for $M=3$ under $N$ expansion}
        \label{fig_m3exp1sa_n}
\end{figure}

\begin{figure}[ht]
\centering
\includegraphics[trim= 105 0 95 0,clip, width=1\linewidth]{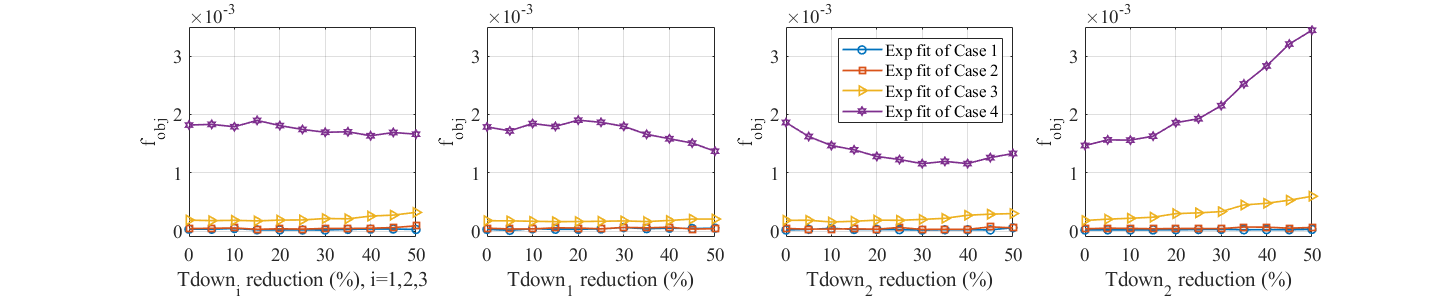}
\caption{$f_{obj}$ resulting from the estimated exponential parameters for $M=3$ under $T_{down}$ reduction}
\label{fig1_g3obj_td}
\end{figure}

\begin{figure}[ht]
     \centering
     \subfloat[Estimation errors of $PR$ resulting from estimated exponential parameters under $T_{down}$ reduction]
         {\includegraphics[trim= 90 0 90 0,clip, width=1\linewidth]{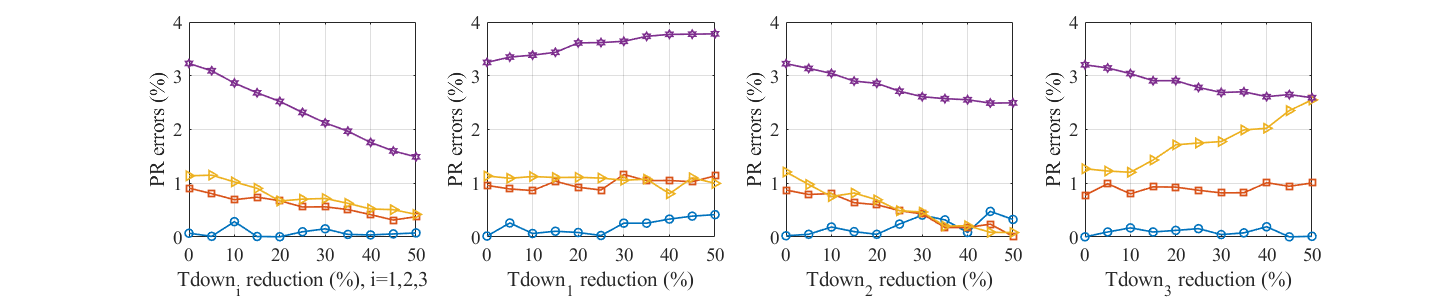}
         \label{fig1_exp1sa_pr_td}}\\
    \subfloat[Estimation errors of $WIP_{avg}$ resulting from estimated exponential parameters under $T_{down}$ reduction]
         {\includegraphics[trim= 90 0 90 0,clip, width=1\linewidth]{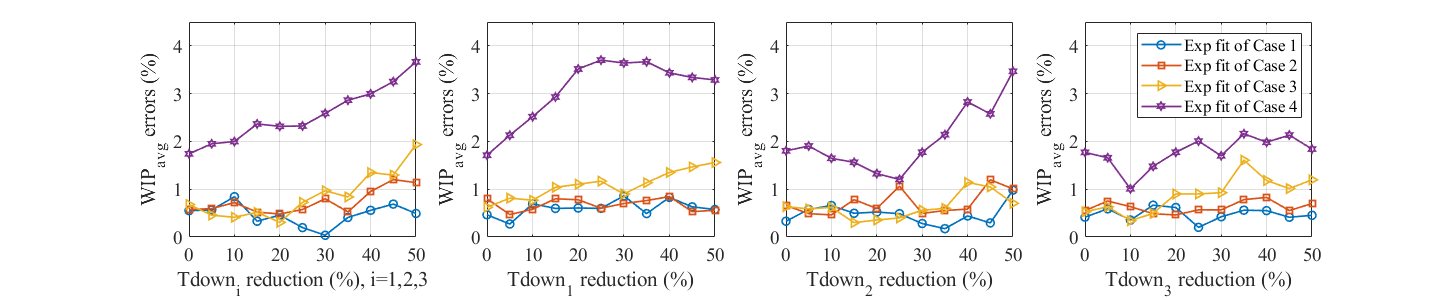}
         \label{fig1_exp1sa_wipa_td}}     
         
\caption{Performance metrics errors resulting from estimated exponential parameters for $M=3$ under $T_{down}$ reduction}
\label{fig_m3exp1sa_td}
\end{figure}

Similar to subsection \ref{par_sen_p}, we investigate the estimated performance metrics under buffer expansion and downtime reduction. For case 4, neither NN-based nor simulated $f_{obj}$ is below the threshold of valid estimation ($10^{-4}$), and we obtain the errors of corresponding performance metrics are much higher than those of the other 3 cases. Definitely, with the changes of $N$'s or $T_{down}$'s, these estimation errors still stay high, as shown in Figure \ref{fig1_g3obj_n} and Figure \ref{fig1_g3obj_td}. 
Moreover, from Figure \ref{fig_m3exp1sa_n} and Figure \ref{fig_m3exp1sa_td} for three-machine cases,
when the buffer capacities are doubled or the downtimes are reduced to half, the maximum of $\epsilon_{PR}$ and $\epsilon_{WIP_j}$ increase to about 7\%. These estimation errors are even higher in five-machine cases. Absolutely,
case 4 is not recommended to fit as an exponential model. 
Thus, for five-machine cases, we exclude case 4 and only plot the objective function values under those improvement scenarios resulting from the exponential fits of cases 1, 2, and 3.

\end{document}